\documentclass[journal=ancac3,manuscript=article]{achemso}

\usepackage{chemformula} 
\usepackage[T1]{fontenc} 
\usepackage{siunitx}
\usepackage{hyperref}
\usepackage{amsmath}
\usepackage{multirow}
\newcommand*{\citen}[1]{%
  \begingroup
    \romannumeral-`\x 
    \setcitestyle{numbers}%
    \cite{#1}%
  \endgroup   
}

\author{Stephen A Church}
\email{stephen.church@manchester.ac.uk}
\author{Nikesh Patel} 
\author{Ruqaiya Al-Abri}
\author{Nawal Al-Amairi}
\affiliation[the University of Manchester]
{Department of Physics and Astronomy and Photon Science Institute, the University of Manchester, Manchester M13 9PL, United Kingdom}
\author{Yunyan Zhang}
\affiliation[Zhejiang University]
{School of Micro-Nano Electronics, Zhejiang University, Hangzhou, Zhejiang 311200, China}
\author{Huiyun Liu}
\affiliation[UCL]
{Department of Electronic and Electrical Engineering, UCL, Malet Pl, London, WC1E 7JE, United Kingdom}
\author{Patrick Parkinson}
\email{patrick.parkinson@manchester.ac.uk}
\affiliation[the University of Manchester]
{Department of Physics and Astronomy and Photon Science Institute, the University of Manchester, Manchester M13 9PL, United Kingdom}

\title[Holistic nanowire laser characterization as a route to optimal design]
  {Holistic nanowire laser characterization as a route to optimal design}


\begin{document}

\begin{abstract}

Nanowire lasers are sought for near-field and on-chip photonic applications as they provide integrable, coherent and monochromatic radiation: the functional performance (threshold and wavelength) is dependent on both the opto-electronic and crystallographic properties of each nanowire. However, scalable bottom-up manufacturing techniques often suffer from inter-nanowire variation, leading to differences in yield and performance between individual nanowires. Establishing the relationship between manufacturing controls, geometric and material properties and the lasing performance is a crucial step towards optimisation, however, this is challenging to achieve due to the interdependance of such properties. Here, a high-throughput correlative approach is presented to characterise over 5000 individual GaAsP/GaAs multiple quantum well nanowire lasers. Fitting the spontaneous emission provides the threshold carrier density, while coherence length measurements determine the end-facet reflectivity. The performance is intrinsically related to the width of a single quantum well due to quantum confinement and bandfilling effects. Unexpectedly, there is no strong relationship between the properties of the lasing cavity and the threshold: instead the threshold is negatively correlated with the non-radiative recombination lifetime of the carriers. This approach therefore provides an optimisation strategy that is not accessible through small-scale studies.

\end{abstract}


\section{Introduction}
Semiconductor nanowires (NWs) can be designed to act as both a gain medium and a Fabry-Perot cavity, due to the high reflectivity of their end-facets~\cite{Maslov2003}, facilitating room temperature lasing with appropriate optical pumping~\cite{Jevtics2017}. These structures can provide monochromatic and coherent light sources for photonic circuits~\cite{Jevtics2017} and sensing applications~\cite{Wu2018d} and can advantageously be grown directly on silicon substrates~\cite{Mayer2016a} and waveguides~\cite{Bermudez-Urena2017b}. 

The performance of NW lasers (NWLs) can be characterised by the lasing threshold and the optical modal structure, which must be optimized to the specific application. The NWL platform is compatible with a wide variety of material systems and is therefore adaptable: by selecting the appropriate gain medium material, lasing wavelengths ranging from the UV (ZnO~\cite{Johnson2003}, GaN~\cite{Johnson2002,Grade2005}) to the visible (CdS~\cite{Duan2003}, perovskites~\cite{Wang2018}) and the IR (GaAs~\cite{Saxena2013}, InP~\cite{Gao2014}, CdSe~\cite{Xiao2011}) have been demonstrated. Further control of the lasing wavelength can be achieved by changing the material composition~\cite{Wang2018}, or by incorporating a heterostructure~\cite{Qian2008e}. Careful design of these can also reduce the lasing threshold by increasing the spatial overlap between the gain medium and the lasing mode~\cite{Saxena2016}. However, variation in one or more of the properties such as the NW size, material quality or cavity can lead to significant changes in their performance~\cite{Church2022a}. It is therefore essential to develop an experimental understanding of how each of these parameters impact the performance to facilitate future fabrication improvements.

The lasing cavity is expected to determine the modal structure of the lasing emission~\cite{Grade2005}. For Fabry-Perot cavities, this is defined by the end facets of the NW so that the NW length determines the longitudinal modes which can lase~\cite{Aman2021}. Mode confinement must also be considered, which is influenced by the NW width, the lasing wavelength, the lasing mode and the refractive index difference between the NW and the surroundings~\cite{Zimmler2010}. 

The reflectivity of the end facets impacts both the degree of optical feedback in the laser and the output coupling. For in-plane NWLs the native reflectivity of the end facets is enhanced above the Fresnel reflection coefficients~\cite{Maslov2003} and is often sufficient for lasing~\cite{Alanis2017}. This can be further enhanced using reflective layers designed for the lasing wavelength~\cite{Zhang2021}. This reflectivity is strongly coupled to the mode confinement in the NW, and theoretical studies have shown that these relationships can be complex~\cite{Maslov2003}. This is because optical far-field approximations are not valid when considering the behaviours of a sub-wavelength cavity - and as a result these effects are challenging to investigate experimentally.

For both vapor-liquid solid (VLS) and selective area growth (SAE) NWs there is an additional complication: since the growth is driven by thermodynamic processes, the NWs demonstrate variation in material and cavity properties that influence the lasing performance~\cite{Al-Abri2021}. It is therefore difficult to vary a single property in isolation, which makes any systematic study of these effects a challenge.

We utilise advancements in laboratory automation to explore the complex, multidimensional parameter space created by inter-NW heterogeneity~\cite{Church2022}. This methodology individually characterises the gain and cavity properties of thousands of NWLs using a suite of optical characterisation techniques. This includes power-dependent photoluminescence spectroscopy, to assess the lasing wavelength and thresholds, time-correlated single photon counting, to determine the carrier lifetimes, waveguiding studies to investigate distributed losses~\cite{Barrelet2004} and interferometry to determine the laser coherence length and end-facet reflectivities~\cite{Skalsky2020}. Correlating the results from these independent measurements provides unique insights into the lasing properties of NWs that would otherwise be inaccessible.

We have applied this approach to 5195 GaAs/GaAsP NWLs that have three, highly-strained quantum wells (QWs) in a core-shell structure, and threshold fluences as low as \SI{6}{\micro\joule\per\centi\meter\squared} at room temperature~\cite{Zhang2019j,Skalsky2020}. We correlate the properties of the cavity and gain medium to the threshold carrier density in the QWs and demonstrate that it is the properties of the QWs that limit the NWL performance. The approach is modular and scalable by design, and therefore suitable for characterisation of other NWL material systems, whilst being widely applicable to emerging opto-electronic materials.

\section{Results and Discussion}

For NWLs, the lasing threshold gain is driven by the interplay between gain and loss mechanisms in the Fabry-P\'erot cavity. The amount of gain is proposed to be determined by both the design and material quality of the active region~\cite{Saxena2015b}. The losses, on the other hand, will likely be affected by the properties of the lasing cavity~\cite{Church2022a}, which is predominantly determined by the reflectivity of the end facets~\cite{Maslov2003}, along with distributed losses that may be driven by reabsorption of lasing radiation~\cite{Alanis2019OpticalLasing} and waveguide losses.

However, the magnitude of these relationships has not been established. To investigate the factors that limit the lasing performance of the NWs, it is necessary to characterise all of these gain and cavity properties of each NW. Specifically, we use this to identify the strength of the correlation between performance and cavity parameters, and performance and material parameters: this can be used to determine an optimisation strategy. This approach was initially applied to single NWs, before being scaled up to a large number of individual NWs to understand how these different properties influence each other.

\subsection{Single nanowire results}

A population of 5195 GaAs/GaAsP multiple quantum well (QW) NWs, grown using a previously described recipe~\cite{Skalsky2020}, were removed from their growth substrate into solution and drop-cast onto a Si substrate to study their optical properties.

\begin{figure*}
    \centering
    \includegraphics[width = 0.95\linewidth]{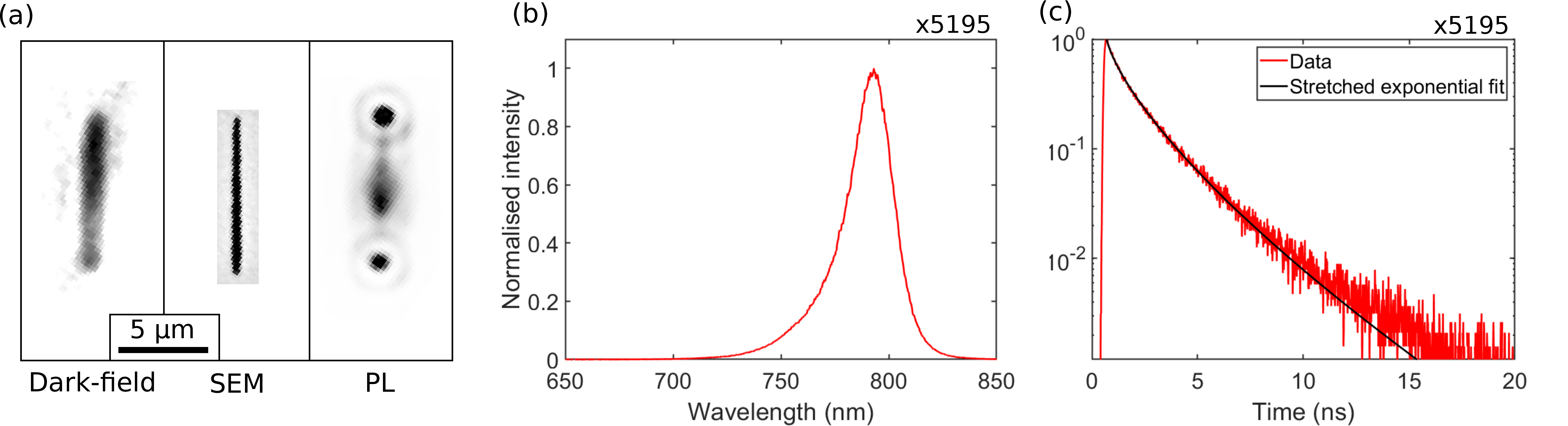}
    \caption{Imaging and spectroscopic characterisation for NW A. The numbers (x----) provide the number of NWs that a successful experiment and analysis was performed. (a) A dark-field optical image, SEM image and PL-emission image of the same NW. (b) A PL spectrum, recorded under low power excitation with a CW laser. (c) A PL time decay, also measured with a low fluence: this has been fit with a stretched exponential decay, defined in Equation~\ref{equ:stretched}. }
    \label{fig:1}
\end{figure*}

A dark field optical microscopy image was recorded for each NW, an example for a specific NW (labelled A) is shown in Figure~\ref{fig:1}(a): image analysis was performed to determine the width and length of each NW. The measured NW width was limited by the resolution of the microscope; therefore supplementary scanning electron microscopy (SEM) was also performed, providing a higher-resolution image, also shown in Figure~\ref{fig:1}(a): SEM and optical images were matched for a subset of 2492 NWs. The mean NW width, and standard deviation (SD), was \SI{0.77(25)}{\micro\meter} and the mean length was \SI{16(4)}{\micro\meter} . Further details of the image comparison and statistics are given in the supporting information (S.I).

Low fluence photoluminescence (PL) spectroscopy was performed by exciting each NW using a frequency-doubled DPSS continuous wave \SI{532}{\nano\meter} laser, focused to a spot diameter of \SI{3}{\micro\meter} with a power of \SI{0.5}{\milli\watt} at the sample. The spatial emission profile of the NW was measuring using imaging (Figure~\ref{fig:1}(a)), and the PL spectrum was also measured: an example spectrum for NW A is shown in Figure~\ref{fig:1}(b). Under these low-power excitation conditions the spectral shape is independent of the excitation power density. There is also minimal axial spectral variation along the NW. Therefore, for below-threshold studies on these NWs, this spectral measurement at a single point provides a result that is both reproducible and representative of the entire NW. 

Time correlated single-photon counting measurements (TCSPC) were performed far below lasing threshold using \SI{633}{\nano\meter}, \SI{200}{\femto\second} excitation pulses from an OPA system, that were attenuated and defocussed at the sample position to achieve an excitation fluence of approximately \SI{15}{\micro\joule\per\centi\meter\squared} across a uniform spot of \SI{63}{\micro\meter} diameter. An example of the resultant PL time decay on NW A is shown in Figure~\ref{fig:1}(c). The recombination lifetime was was fit by a stretched exponential model defined by Equation~\ref{equ:stretched}, with a power factor of \SI{0.71}: this suggests that there is a degree of disorder in the QWs~\cite{Dyer2021a}. The lifetime was \SI{1.0}{\nano\second} for this NW. Under these conditions, both radiative and non-radiative Shockley-Reed-Hall (SRH) recombination are likely to contribute to these dynamics, which is explored further in the S.I. An estimated IQE of \SI{39}{\percent} was determined using this approach. Further analysis of the IQE and the power factor is provided in the S.I.

\begin{figure*}
    \centering
    \includegraphics[width = 0.95\linewidth]{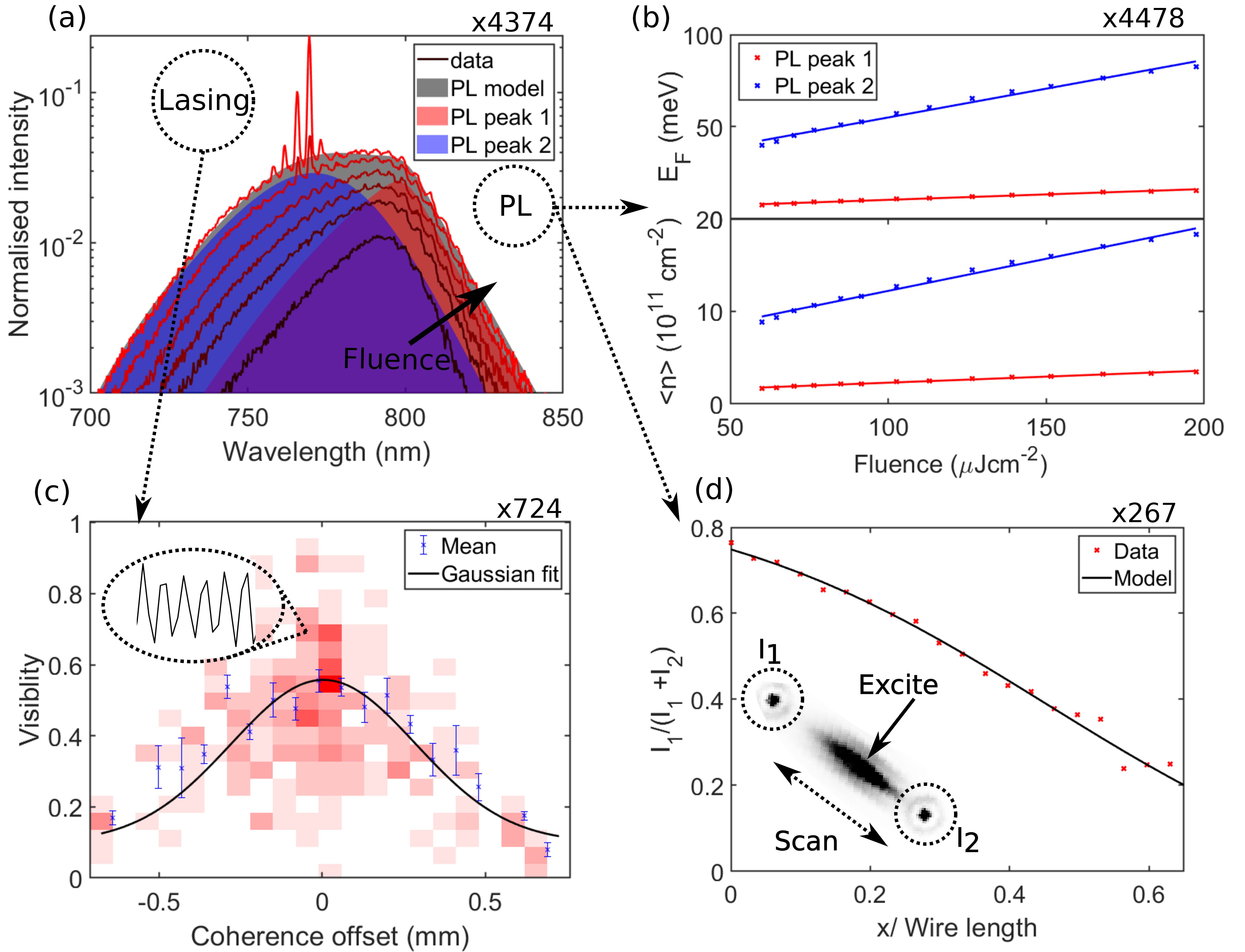}
    \caption{Characterisation of the lasing properties for NW B. The numbers (x----) provide the number of NWs that a successful experiment and analysis was performed. (a) Emission spectra showing PL and lasing emission with excitation fluence varying from \SI{19}{\micro\joule\per\centi\meter\squared\per pulse} to \SI{270}{\micro\joule\per\centi\meter\squared\per pulse}. An LSW fit is shown applied to the highest fluence PL spectrum, demonstrating the contribution from two PL peaks. (b) The variation of Fermi energy and carrier density with excitation fluence in two different QWs, extracted from LSW fits to the PL spectra. A linear fit to the data has been shown. (c) A 2D histogram showing how the visibility of the interference fringes changes with distance from the zero path length position. The median and SD values for each horizontal bin are given in blue as a guide for the eye. A Gaussian fit has been applied to the visibility data, to extract the coherence length. Inset is a section of the raw interferogram, showing the interference fringes. (d) A plot of the intensity of light coupling out of the NW end-facets versus position of the excitation spot. This has been fit with a model, defined by Equation~\ref{equ:distributed}, to extract a value for the distributed losses.}
    \label{fig:2}
\end{figure*}

The lasing properties of each NW were characterised by recording the emission spectrum at different excitation fluences. For these experiments, the same pulsed laser system was used to achieved a uniform spot of \SI{73(1)}{\micro\meter} diameter with fluences up to \SI{2000}{\micro\joule\per\centi\meter\squared}. Critically, this wavelength avoids exciting the direct bandgap of the QW barriers~\cite{Skalsky2020}, and so avoids excessive heating from carrier cooling and complications due to carrier capture from the barriers. An example of these spectra for a second wire, NW B. is shown in Figure~\ref{fig:2}(a), which shows spectra for fluences between \SI{19}{\micro\joule\per\centi\meter\squared\per pulse} and \SI{270}{\micro\joule\per\centi\meter\squared\per pulse}. At low excitation fluences, only the broad spontaneous PL emission can be observed. As the fluence is increased above a threshold, a sharp lasing emission peak with a full-width-half-maximum (FWHM) of approximately \SI{1.5}{\nano\meter} appears in the spectrum.

Analysis of power-dependent photoluminescence spectrum provides the primary lasing wavelength, the threshold and the inter-modal spacing. For NWLs where longitudinal mode structure is expected to dominate, this can be further used to determined the effective cavity length. An example of the threshold and length analysis is provided in the S.I. For NW B, the threshold is \SI{230}{\micro\joule\per\centi\meter\squared\per pulse} and the wavelength is \SI{773}{\nano\meter}, and this approach was successfully applied to 4460 (\SI{84}{\percent}) of the NWs. The other NWs could not be analysed automatically due to the emergence of multiple peaks. An additional filtering step was applied to these results to ensure that only threshold values with a low uncertainty were considered for correlation studies. This excluded an additional \SI{2}{\percent} of wires from further analysis, and a similar approach was applied for the other aspects of this study.

The spectra in Figure~\ref{fig:2}(a) also contain a broad photoluminescence emission band, which contains two emission peaks, peak 1 at longer wavelengths, and peak 2 at shorter wavelengths. These peaks were fit using a Lasher-Stern-Wurfel (LSW) model~\cite{Fadaly2020}, for two-dimensional states defined by Equation~\ref{equ:PL_fit} in the experimental section. The PL transition energy, and $\sigma$, which is a measure of the symmetrical disorder in the QW~\cite{Alanis2019OpticalLasing}, were extracted from the fitting results. The Urbach energy is another fitting parameter, that is discussed in the S.I. For NW B, this yielded transition energies of \SI{1.551}{\electronvolt} and \SI{1.591}{\electronvolt}. $\sigma$ was found to be \SI{15}{\milli\electronvolt} and \SI{14}{\milli\electronvolt} for these peaks, indicating a similar degree of disorder in both peaks. We attribute peak 1 to carrier recombination in a QW, and we consider several candidates for peak 2.

Firstly, peak 2 may originate from carrier recombination in different QWs in the core-shell structure. This is supported by considering the band structure of the three QWs in the NW. Due to the close proximity to the NW core, which has a different alloy composition, the inner-most QW has an asymmetrical barrier~\cite{Skalsky2020}. One-dimensional numerical solutions of the Schr\"odinger Equation (SE), shown in the S.I., demonstrate that this effect can increase the E\textsubscript{11} ground state transition energy by \SI{15}{\milli\electronvolt} when compared with the other QWs.

As an alternative explanation, the shorter wavelength recombination peak may be a product of bandfilling effects in the same QW at high fluences. This could result in carrier recombination through the E\textsubscript{22} transition in the same QW. Using the SE solutions, this results in a transition energy that is up to \SI{350}{\milli\electronvolt} higher than the ground state, for realistic QW widths. Alternatively, carriers could be overflowing from the QW into the indirect X state in the QW barriers, which results in a transition energy of \SI{1.98}{\electronvolt}~\cite{Skalsky2020}. These transition energies are both significantly larger than those observed in the PL spectra.

The bandfilling effects in these QWs are directly linked to the FWHM of the emission. A global LSW fit was applied to the full set of fluence-dependent PL spectra in Figure~\ref{fig:2}(a): properties such as the bandgap and $\sigma$ were kept constant with fluence, whilst allowing the Fermi energy and peak amplitude to vary. To minimise the total number of fitting parameters, it was assumed that the Fermi energy and amplitude of each peak were related by a constant factor that was independent of fluence. The Fermi energy at each excitation fluence for NW B is shown in Fig~\ref{fig:2}(b). As this experiment used pulsed excitation, this is the time-averaged $E_F$, or $<E_F>$, which is weighted by the instantaneous intensity of the PL, $I_{PL}(t)$ and is hence skewed towards early times. This is described by:
\begin{equation}
\label{equ:EF_av}
    <E_F> = \frac{E_F(t) \times I_{PL}(t)} {\int_{0}^{\text{\SI{5}{\micro\second}}} I_{PL}(t) \,dt} ,
\end{equation}
integrating over the laser repetition rate of \SI{5}{\micro\second}. At fluences below \SI{50}{\micro\joule\per\centi\meter\squared} the Fermi energy cannot be extracted reliably, due to large fitting errors. However, above \SI{50}{\micro\joule\per\centi\meter\squared}, and at threshold, a linear correlation is observed between excitation fluence and Fermi energy. This suggests that we only observe emission from E\textsubscript{11} transitions in the QWs. The linear fit does not pass through the origin, and so $E_F$ must vary super-linearly at low fluences. This is suggestive of a reduced density of states at low energy due to the impact of disorder on the bandstructure.

A suitable fit was achieved whilst maintaining a constant relative amplitude of the two PL peaks. Therefore, in this fluence regime, there is no significant change in the relative occupation of the states associated with each peak and there is no significant carrier transfer between the states. As a result, it is unlikely that peak 2 originates either from the E\textsubscript{22} transition or from the QW barrier states. We therefore attribute peak 2 to the E\textsubscript{11} transition in the QW closest to the NW core (QW-core), which has a higher transition energy than the other QWs (QW-shell). The zero strain equivalent (ZSE) widths of these QWs were estimated using one dimensional numerical solutions of the SE, and are given in Figure~\ref{fig:3}(b). These calculations do not consider the impact of higher order effects, such as strain in the QW, more details of which are provided in the S.I. Strain is likely to have an impact on the band alignment in the QWs, and therefore the transition energies, but, crucially, the variation of transition energies will remain comparable for small-magnitude strains. Therefore, any variation in the QW width will have a similar effect on the ZSE width. This approach gave a ZSE width of \SI{3.8}{\nano\meter} for the QW-core and \SI{4.6}{\nano\meter} for the QW-shell.

Our model provides a route to study each emission in isolation: As shown in Fig~\ref{fig:2}(a), the emission from the QW-core (peak 2) has a larger FWHM than the QW-shell (peak 1). The LSW model therefore calculates a larger Fermi energy: Fig~\ref{fig:2}(b) shows that that the Fermi energy in the QW-core is around double that of the QW-shell. This is the result of a higher carrier density in the core-QW. COMSOL simulations in the S.I. suggest that the absorption of photons in each QW is expected to be guided by the spatial overlap of the QWs with the electric fields of the excitation.

It is typical for lasing thresholds to be reported in units of fluence (\si{\micro\joule\per\centi\meter\squared})~\cite{Zhu2015,Alanis2017}. However, using this approach, it is difficult to compare results from different NWs due to variation in experimental conditions. For example, the threshold fluence can be highly dependent on the excitation wavelength~\cite{Skalsky2020}. It is also expected that substrates with different refractive indices will modify the threshold by changing the cavity properties~\cite{Saxena2015b}. Additionally, differences in absorption and carrier dynamics may also cause changes in the threshold: this effect is explored by finite difference time domain simulations in the S.I. To remove the resultant ambiguity, it is necessary to move to a more direct measurement of the laser threshold. This can be achieved by considering the threshold areal carrier density within the QWs. 

Our bandfilling analysis provides a route to determine the averaged carrier density as it is related to the Fermi energy. Using Equation~\ref{equ:carr_dens}, defined in the methods section, the time-averaged carrier density in the QWs, $<n>$, was calculated. To convert this to the initial carrier density, $n_0$, the PL time decays were also measured close to the lasing threshold (at 0.6 P\textsubscript{th}) under the same excitation conditions. For NW B, this provided a threshold $n_0$ of \SI{3.9E15}{carriers \per\centi\meter\squared} for the QW-core. This analysis is discussed in detail in the experimental methods, and lifetime statistics are shown in the S.I.

The intensity reflectivities of the end-facets of the NW were determined using an interferometric approach based on ref~\cite{Skalsky2020}. The interferogram of the lasing emission above threshold was measured and the fringe visibility was calculated by measuring the peak-to-peak amplitude across the interferogram, using Equation~\ref{equ:visibility}: Figure~\ref{fig:2}(c) shows how the fringe visibility varies with distance around the zero path-length position. The scatter around the fitted Gaussian is symmetrical in nature, and originates from random error in the sampling position of each point in the interferogram, which is a consequence of the fast stage speeds required for high-throughput measurement. The FWHM of the Gaussian is equal to the coherence length of the emission, which is \SI{0.7}{\milli\meter} for NW B. This coherence length is much larger than the NW length (\SI{20}{\micro\meter}), and corresponds to a lasing Q factor of 900. 

Using Equation~\ref{equ:reflect}, in the methods section, the coherence length was used to calculate the intensity reflectivity of the cavity, described as the geometric mean of the end-facet reflectivity. This had a value of 0.56 for NW B. This reflectivity is larger than the far-field reflectivity calculated using the Fresnel reflection coefficients (0.30), as expected from theoretical calculations~\cite{Maslov2003}. Fluorescence imaging was used to determine the reflectivity of each facet separately: this analysis is shown in the S.I. The reflectivity measurements were performed for 1737 NWs, successfully finding the reflectivity of 724 NWs. The roughly \SI{60}{\percent} failure rate in our analysis was due to a relatively low collected intensity of laser radiation from each NW, which resulted in large uncertainties in the coherence length and reflectivity values.

The distributed losses in the lasing cavity were estimated by imaging the NW emission when exciting with a \SI{2}{\micro\meter}-diameter spot from a \SI{0.5}{\milli\watt} CW HeNe laser. The NW orientation angle was determined through automatic analysis of the images: this enabled, as depicted in the inset of Figure 2(d), the excitation spot to be automatically scanned along the NW length and the emission intensity at the end facets to be recorded. This scanning approach can be scaled up to study 100's of individual NWs without the need for further human input. The change in intensity of light emission from each end-facet was measured to produce the data shown in Figure~\ref{fig:2}(d). The intensity of emission from a facet drops approximately exponentially with distance between the excitation point and the facet. Using the Beer-Lambert Law, the distributed losses, $\alpha$, were found from the ratio of the integrated intensity from each facet, $I_{1,2}$. This is given by Equation~\ref{equ:distributed}:
\begin{equation}
\label{equ:distributed}
    \frac{I_1}{I_1 + I_2} = A + \left(   1 + B \left( \frac{1-R_2}{1-R_1} \right)  \frac{\text{exp}(-\alpha(L-x))}{\text{exp}(-\alpha x)} \right)^{-1}
\end{equation}
where $R_{1,2}$ is the reflectivity of each end facet, $x$ is the distance from facet 1, $A$ is a term accounting for a background baseline in the measurement and $B$ accounts for additional losses. $\alpha$ was \SI{1300(100)}{\per\centi\meter} for NW B. The losses were measured for a subset of 267 NWs and they may be due to scattering or coupling from the cavity side-wall, or reabsorption~\cite{Alanis2019OpticalLasing}.

\subsection{Population studies}

\begin{figure*}
    \centering
    \includegraphics[width = 0.95\linewidth]{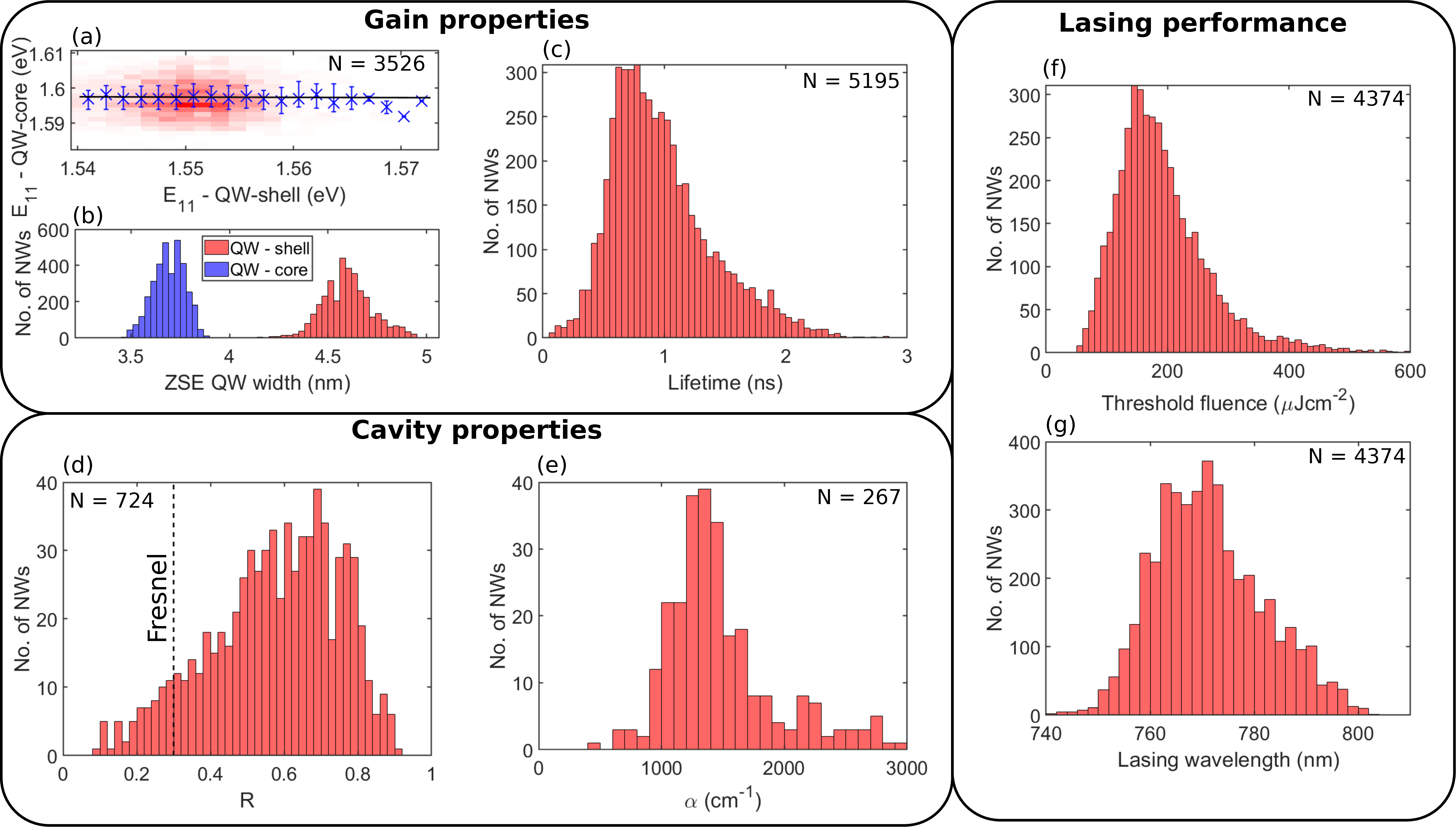}
    \caption{Distributions of the parameters describing the gain medium, cavity properties and lasing performance, across the NW population. (a) A two-dimensional histogram showing the lack of correlation between the E\textsubscript{11} transition energies of each QW. The blue crossed and error bars are the median and the inter-quartile range values for each bin. (b) The estimated ZSE width of two QWs in the core/shell structure, \SI{4.6(1)}{\nano\meter} and \SI{3.7(1)}{\nano\meter}. (c) The carrier recombination lifetime \SI{0.9(4)}{\nano\second}. (d) The cavity reflectivity \SI{0.6(2)}. (e) The distributed losses \SI{1400(400)}{\per\centi\meter}. (f) The threshold fluence \SI{180(80)}{\micro\joule\per\centi\meter\squared\per pulse}. (g) The initial lasing wavelength \SI{770(10)}{\nano\meter}.}
    \label{fig:3}
\end{figure*}

To decouple the impact of lasing cavity and material gain from the threshold, the above investigations were repeated on a large number of individual NWs. As a result, distributions for the previously determined parameters were obtained, which demonstrate the variation across the sample population. These results are summarised by the histograms in Figure~\ref{fig:3}. 

Analysis of the PL spectra extracted the transition energies associated with each PL emission peak for 3526 NWs (Figure~\ref{fig:3}(a)), which demonstrates that there is no correlation between these transition energies. This observation, that could not be made from a single-NW measurement, provides further evidence that the shorter wavelength peak does not originate from a QW that is independent from the longer wavelength peak QW.

Additional measurements determined the lifetimes of 5195 NWs (Figure~\ref{fig:3}(c)), interferometry was used to find the reflectivity for 724 NWs (Figure~\ref{fig:3}(d)), the distributed losses were characterised for 267 NWs ((Figure~\ref{fig:3}(e)) and the fluence dependence study found the lasing thresholds and wavelengths for 4374 NWs (Figs.~\ref{fig:3}(f) and (g)). The "best in class" NW had a threshold fluence of \SI{51}{\micro\joule\per\centi\meter\squared\per pulse} at a wavelength of \SI{793}{\nano\meter}.

There is variation in all of the reported parameters, but this is of particular note for the intensity reflectivity, which varies between 0.1 and 0.9. This variation is significantly larger than the uncertainty in parameters that are assumed to be constant, such as the refractive index (which has a \SI{15}{\percent} uncertainty, see the S.I. for more information). Furthermore, the reflectivity does not correlate with any other cavity or gain parameters, and so the variation is likely to be caused by other factors. This could either be related to variation in roughness and faceting of the end of the NW that is broken during the process of transferring the NW from their native substrate~\cite{Alanis2019a}, or due to irregular morphology at the NW tip, due to an increase in defects~\cite{Zhang2019a}. It has been suggested that deviation from the ideal planar morphology could reduce the reflectivity by up to a factor of 5~\cite{Saxena2016}, which may account for the majority of the observed variation. 

The reflectivity of \SI{91}{\percent} of the NWs is larger than the Fresnel coefficient. This result is compatible with early simulations by Maslov and Ning~\cite{Maslov2003}, who demonstrated that, in the regime of strong waveguiding, the intensity reflectivity of the end facets can be up to three times higher than an infinite planar surface.

The distributions represent a coupled, multidimensional dataset of the NW population. This was analysed using a multiple linear regression model, considering the impact of changes in the lasing cavity and gain medium on the lasing threshold and wavelength. The model parameters are summarised in Table~\ref{table:mlrm}, and will be discussed further in this section.

\begin{table*}[]
\small
\centering
\caption{Outputs from multiple linear regression models that quantify the relative impact of varying NW properties on three different performance metrics. The variation of the performance metric with each variable is provided as a percentage of the maximum observed variation, and the p-value for each variable is also shown. Variables with a significance within the $3\sigma$ ($p = 0.003$) limit are highlighted in bold. }
\label{table:mlrm}

\begin{tabular}{clcccccc}
\hline
\multicolumn{1}{l}{}    &                       & \multicolumn{2}{c}{\begin{tabular}[c]{@{}c@{}}Lasing \\ wavelength\end{tabular}} &  \multicolumn{2}{c}{\begin{tabular}[c]{@{}c@{}}Threshold \\ carrier density\end{tabular}} \\ \hline
\multicolumn{1}{c}{}    & Variable              & \multicolumn{1}{c}{Variation (\%)}         & \multicolumn{1}{c}{p-value}         & \multicolumn{1}{l}{Variation (\%)}         & \multicolumn{1}{c}{p-value}                  \\ \hline
\multirow{4}{*}{Gain}   

                        & Lifetime              & -4                                         & 0.50                                                        & \textbf{-36}                                     & \textbf{$\sim$0}                         \\ 
                        
                        & $\sigma$             & -11                                         & 0.11                                                        & \textbf{-18}                                     & \textbf{$\sim$0}                         \\

& QW-shell ZSE width            & -1                                         & 0.95                                                   & -3      & 0.69    \\& 
QW-core ZSE width            & \textbf{48}                                & \textbf{$\sim$0}                                    & \textbf{-26}     & \textbf{$\sim$0}                         \\ \hline
\multirow{4}{*}{Cavity} & NW width              & 3                                          & 0.61                                                         & 3                                                & 0.52                                     \\
                        & NW length             & 0                                          & 0.90                                              & \textbf{-13}                                     & \textbf{$\sim$0}                  \\
                        & $\alpha$ & -10                                         & 0.47                                                           & -14                                             & 0.19                                     \\
                        & Reflectivity          & 0                                          & 0.85                                                        & -3                                               & 0.21                                     \\ \hline
\end{tabular}
\end{table*}

\begin{figure*}
    \centering
    \includegraphics[width = 0.95\linewidth]{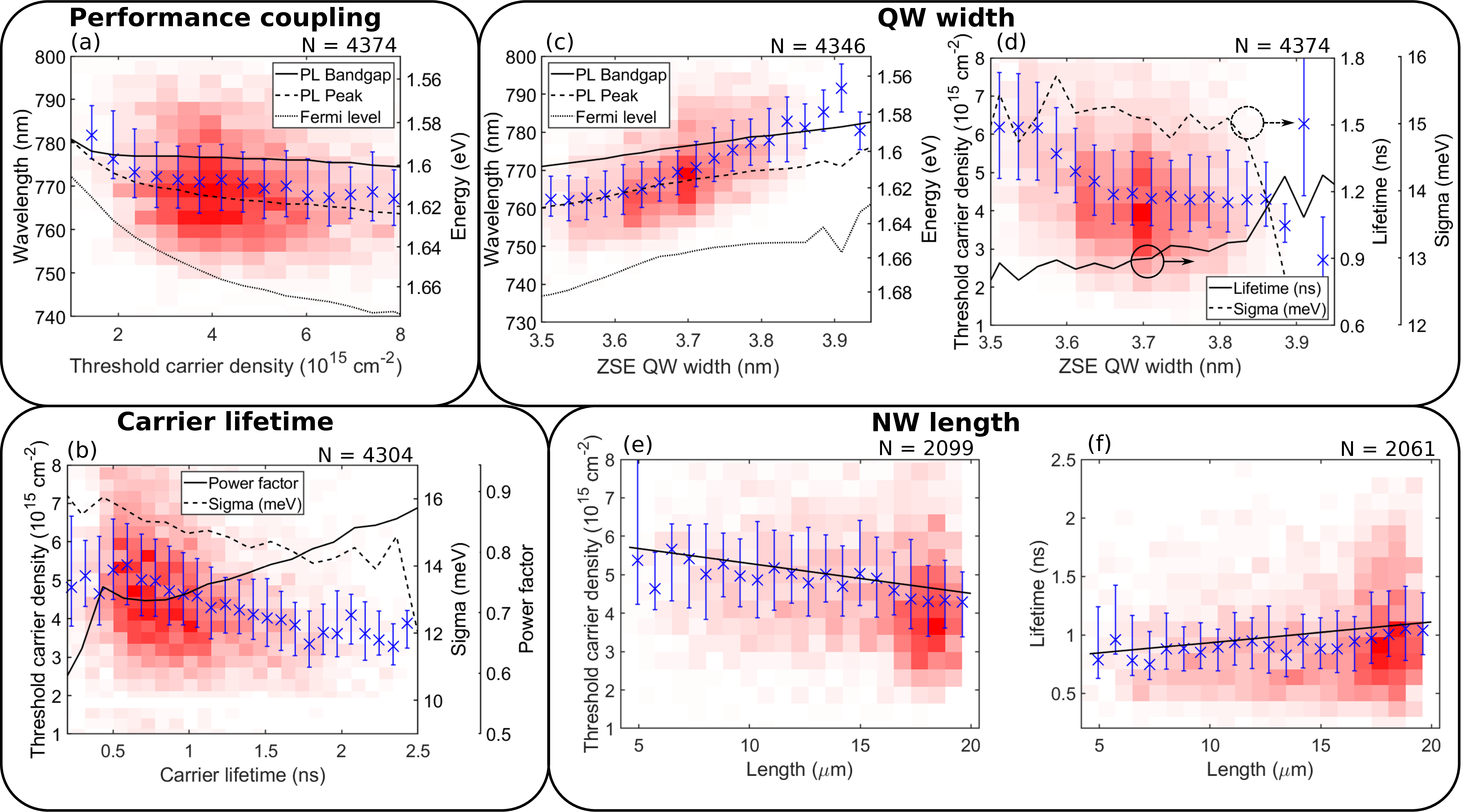}
    \caption{Correlations between different parameters observed across the population. Each plot contains data from the NWs where successful fitting of the independent measurements has been performed: data-points represent median values in each horizontal bin, with the error bars indicating the inter-quartile range. (a) Lasing wavelength vs threshold carrier density for 4374 NWs. The plot also includes the median wavelengths corresponding to the PL bandgap, the peak in the LSW PL spectrum and the Fermi level. (b) Threshold carrier density vs carrier lifetime for 4304 NWs, a linear fit through the data is shown as a guide to the eye. The plot also includes the median $\sigma$, determined from the PL fitting, and the median power factor, taken from the stretched exponential fits to the PL decays at low fluence. (c) Lasing wavelength vs QW ZSE width for 4346 NWs. The plot also includes the median wavelengths corresponding to the PL bandgap, the peak in the LSW PL spectrum and the Fermi level. (d) Threshold carrier density vs QW ZSE width for 4374 NWs. The plot also includes the median carrier lifetimes and $\sigma$ (disorder) values from the PL fitting (see Methods). (e) Threshold carrier density vs NW length for 2099 NWs, a linear fit through the data is shown as a guide to the eye. (f) Threshold carrier density vs carrier lifetime for 2061 NWs, a linear fit through the data is shown as a guide to the eye.}
    \label{fig:4}
\end{figure*}

Firstly, the lasing performance metrics are strongly coupled together. This is demonstrated in Figure~\ref{fig:4}(a) for the lasing wavelength and threshold carrier density, where an order of magnitude increase in the threshold causes a blueshift of the median lasing wavelength by approximately \SI{10}{\nano\meter}. Reabsorption of emitted light can be invoked to explain such behaviour~\cite{Alanis2017}, however the performance of NWs in these previous reports was dominated by reabsorption in the NW core. For the GaAs/GaAsP NWs studied in this paper, the core has a larger bandgap than the QWs and so reabsorption should be reduced, and therefore the observed variation may have a different origin. 

 As demonstrated in Figure~\ref{fig:4}(a), the median Fermi energy also varies between \SI{20}{\milli\electronvolt} and \SI{80}{\milli\electronvolt}. This results in a smaller shift in the peak PL wavelength. This shift is due to an increased degree of bandfilling at higher carrier densities. The median lasing wavelength lies between the bandgap and the Fermi level, and the trend in lasing wavelength matches well with the variation in peak PL wavelength. This suggests that the correlation between lasing wavelength and threshold is a signature of bandfilling effects in the QW. The lasing wavelength is consistently redshifted by \SI{2}{\nano\meter} relative to the PL peak, which suggests that there may be an increased modal overlap with the QWs for longer wavelengths.

The effect of the gain medium is determined by looking at the impact of the QW properties. Table~\ref{table:mlrm} shows that the carrier recombination lifetime in the QWs has no relationship with the lasing wavelength, there is however a strong negative correlation with lasing threshold, which is illustrated in Figure~\ref{fig:4}(b). The measured lifetimes vary between approximately \SI{0.5}{\nano\second} and \SI{2.5}{\nano\second}. We find that under these conditions, non-radiative recombination has a significant impact on the recombination rate (see S.I. for details). A decrease in the non-radiative rate causes a reduction in the threshold. The non-radiative rate is strongly influenced by defects in the QWs or barriers layers~\cite{Vening1993,Zhang2019a}; therefore a higher quality gain medium, with a lower density of defects, results in a lower threshold. By improving the gain medium quality, by, for example, optimising the growth of the unintentionally defective NW tip~\cite{Zhang2019a}, the threshold can be minimised.

$\sigma$ is a measure of the disorder in the QWs, which is extracted from the FWHM of the PL spectra. As shown in Figure~\ref{fig:4}(b), the median value of $\sigma$ varies between \SI{16}{\milli\electronvolt} and \SI{13}{\milli\electronvolt} and is strongly negatively correlated with the the carrier lifetime. This correlation is statistically significant, with a Pearson's r value of -0.18 and a p-value close to zero. Therefore, a NW with a lower defect density also has a smaller $\sigma$, and a reduced degree of disorder. This correlation is expected since defects and opto-electronic disorder typically go hand-in-hand. Table~\ref{table:mlrm} demonstrates that this results in a statistically significant negative correlation between $\sigma$ and the lasing threshold. This trend is compatible by the variation in the power factor, extracted from the stretched exponential fitting of the PL decays, which is smaller for larger values of sigma, and higher thresholds (Figure~\ref{fig:4}(b)). A smaller power factor is the result of a greater deviation from an exponential PL decays, which is another signature of disorder within the system~\cite{Dyer2021a}. More statistics on the power factor can be found in the S.I. 

Intriguingly Table~\ref{table:mlrm} also shows that there is no significant correlation between the ZSE width of the thicker QW-shell and the lasing performance. In contrast, changes in the ZSE width of the thinner QW-core cause changes in all of the performance metrics. This means that it is solely QW-core that is the gain medium in this laser system. This is an important observation that is facilitated by the high-throughput approach and can be used to inform future NWL production by removing the QWs that do not contribute to the gain. 

There is a positive correlation between the QW-core ZSE width and lasing wavelength: this is shown in Figure~\ref{fig:4}(c), and corresponds to a redshift of the median lasing wavelength of approximately \SI{20}{\nano\meter} for an increase in the QW-core ZSE width from \SI{3.5}{\nano\meter} to \SI{3.9}{\nano\meter}. As shown in Figure~\ref{fig:4}(c), we would only expect a redshift of \SI{13}{\nano\meter} in the bandgap over this range due to a reduction in the quantum confinement energy. The observed variation is therefore is a steeper than the trend in the PL bandgap, and it can not be explained purely by a reduction in the quantum confinement energy. 

To explain this additional wavelength shift, we initially consider an increase in the bandfilling effects for narrower QW-cores, due to a reduced number of states. As shown in Figure~\ref{fig:4}(c), the median PL peak shifts by approximately \SI{15}{\nano\meter} across the dataset. Therefore, there is an additional \SI{5}{\nano\meter} redshift that is unaccounted for. We tentatively assign this to a shift in the lasing gain curve due to changes in the mode overlap with the QW-core as the lasing wavelength and QW-core ZSE width change. It is difficult to analytically quantify this effect at this stage, but this observation offers an interesting direction for future studies.

The QW-core ZSE width also has a negative correlation with the lasing threshold, as shown in Figure~\ref{fig:4}(d). The median threshold carrier density decreases from \SI{6.3E15}{\per\centi\meter\squared} to \SI{4.3E15}{\per\centi\meter\squared} for an increase in the QW ZSE width from \SI{3.5}{\nano\meter} to \SI{3.95}{\nano\meter}. This variation can largely be explained by considering how the QW ZSE width is coupled to the QW properties, which is demonstrated in Figure~\ref{fig:4}(d). A narrower QW-core results in increased values of $\sigma$, and therefore increased disorder, which is compatible with previous studies~\cite{Davies2015b}. These narrower QW-cores also have, on average, a faster recombination lifetime: this suggests that there may be a faster non-radiative recombination rate and thus a higher density of defects. The nature of these defects is currently unknown, and is a topic for further study: however, they are not likely to be strain related, as a previous study found no evidence of misfit dislocations at the GaAsP/GaAs interfaces. Despite this uncertainty over the type of defects, it is clear that the quality of the QW-cores can be improved by targeting a wider QW-core during growth, which will lead to lower thresholds.

The effect of the lasing cavity on the performance can be gauged by studying the impact of the NW dimensions. Table~\ref{table:mlrm} demonstrates that the NW width has no statistically significant impact on performance. This is likely to occur for NWLs that are wide in comparison to the lasing wavelength, and has been theoretically predicted for both AlGaAs/GaAs [15] and InP [26] NWLs. The NWs in this study have a mean width of (0.77 ± 0.25) µm, which is significantly larger than the lasing modal wavelength, $\lambda /n = 270$ nm. Furthermore, as the lasing gain originates solely from the QW closest to the NW core, it is reasonable to expect strong confinement of the mode to the lasing cavity and furthermore that the modal refractive index is not significantly affected by the NW width. This conclusion is also consistent with analysis in the S.I. that shows a small uncertainty in the refractive index that does not correlate with NW width.

The NW length also has no impact upon the lasing wavelength, likely because these are an order of magnitude longer than the lasing wavelengths, so any impact of longitudinal mode selection will be minimal. Table~\ref{table:mlrm} shows that there is a slight negative correlation with the threshold, which is shown in Figure\ref{fig:4}(e). A similar trend has been observed previously in AlGaAs/GaAs MQW NWLs~\cite{Alanis2017}, which was attributed to gain in the cavity overcoming the end-facet losses with increasing length. Whilst this is expected for the GaAsP/GaAs NWLs, Table~\ref{table:mlrm} shows that changes in the distributed losses, $\alpha$, do not correlate with the performance, and so this effect will be small with respect to the other observed trends. For these wires, a larger effect is an observed correlation between the NW length and the carrier lifetime, as shown in Figure\ref{fig:4}(f): this means that a longer NW will, on average, have a longer carrier recombination time, a higher quality QW and therefore a lower threshold. This is compatible with a previous transmission electron microscopy study on these NWs that observed the formation of a high density of threading dislocations close to the NW tip~\cite{Zhang2019a}. These dislocations act as non-radiative recombination centres, but as the nominal NW length increases, the relative importance of this recombination pathway will be reduced because the proportion of the gain medium that is defective will be lower. This, in turn, results in a lower threshold.

The cavity reflectivity does not correlate with the wavelength or threshold. Additionally, there is no relationship between the cavity reflectivity and the NW width: this is consistent with the large NW widths and the lack of correlation between NW width and performance.

However, the result for $\alpha$ is surprising, which is expected to impact the lasing performance~\cite{Hirano2005}. However, this result does not preclude the effect $\alpha$ on performance, merely that variation in the gain medium (i.e. carrier lifetime and QW width) has a stronger and limiting impact on the performance. Due to the large number of independent variables and complex inter-dependencies in this data, these conclusions can only be reached through the big-data approach.

\section{Conclusion}

A multi-modal automated optical characterization approach was developed to study the performance of a large number of individual NW lasers, including optical imaging, spectroscopic, single-photon counting and interferometric measurements. When applied to 5195 GaAsP/GaAs NW lasers, this created a multi-dimensional dataset including 18320 individual measurements. Correlations were drawn between structural (e.g. length, width) , material (e.g. bandgap, carrier), cavity (e.g. reflectivity, losses) and functional (e.g. lasing threshold, wavelength) properties to gain a deeper understanding into the processes that govern lasing performance and establish the route to improve this performance.

By analysing the PL lineshape, it was found that the properties of only the QW closest to the NW core correlated with the lasing performance. This both enabled the unambiguous identification of the gain medium, and provided the means to calculate the carrier density in the QW at threshold. 

The performance metrics of threshold carrier density and lasing wavelength were compared with independent measurements of the cavity and gain medium using a multiple linear regression model. This analysis demonstrated that the lasing wavelength was influenced by the lasing threshold, due to bandfilling effects and the QW width, partially due to quantum confinement effects. The width of the QWs also directly influences the disorder and the nonradiative defect density which leads to reduced thresholds for the wider QWs. The dominant factor influencing the lasing threshold was the carrier recombination lifetime. Therefore, to minimise the lasing threshold, further optimisation of the QW is required to minimise the non-radiative recombination rate, which can be achieved by increasing the length of the NWs and the width of the QWs.

This big-data approach to experimental analysis of NWLs can therefore provide conclusions that are not accessible using small-population studies. Crucially, the approach needs minimal \textit{a-priori} information, in this case only requiring knowledge of the refractive index and the core/shell heterostructure and can be widely applied to similar materials systems.


\section{Experimental Section}
\subsubsection{Sample growth}
The NWs reported in this paper were grown to a previously published recipe using self-catalysed molecular beam epitaxy on Si substrates~\cite{Zhang2019j}. The NW core has a nominal width of \SI{80}{\nano\meter} and a composition of GaAs\textsubscript{0.38}P\textsubscript{0.62}. This is followed by three GaAs QWs with \SI{40}{\nano\meter} GaAs\textsubscript{0.47}P\textsubscript{0.53} barriers and a \SI{30}{\nano\meter} Al\textsubscript{0.5}Ga\textsubscript{0.5}As\textsubscript{0.53}P\textsubscript{0.47} passivation layer.

Before characterisation, the NWs from these arrays were transferred onto a Si substrate via ultrasonication and drop casting~\cite{Alanis2019a}. The result was a large population of NWs ($>$ 5000) in the plane of the substrate, which is ideal for detailed optical characterisation.

\subsubsection{Automated microscopy}

To characterise the geometry and the optoelectronic performance of the NWs, the transferred substrate was placed in a bespoke, automated optical microscope. This setup utilised a 20x objective lens with a numerical aperture of 0.75 and a working distance of \SI{1}{\milli\meter}. The objective was vertically mounted on a P.I. V-308 translation stage. Optical imaging was performed using front illumination with a \SI{200}{\milli\meter} focal length tube lens to produce a magnified image with a spatial resolution of \SI{1}{\micro\meter}. Confocal laser excitation was also possible using a series of beamsplitters. 

The NW samples were placed on a P.I. V-738 x-y translation stage, capable of a travel range of \SI{150}{\milli\meter} by \SI{150}{\milli\meter}, with \SI{0.1}{\micro\meter} precision. Initially, a pulse train from an excitation laser was focused onto the substrate to ablate material in order to create markers. These markers were separated by \SI{2}{\milli\meter} and defined an x-y co-ordinate system on the substrate which could be used to uniquely locate each object.

The NWs were located using an automated microscopy procedure. This involved translating the x-y stage to a new region of the substrate, refocusing the objective lens to correct for any slope in the sample and imaging the surface. A machine vision algorithm automatically identified NWs based on their dimensions and appearance. The location, width, length and orientation angle for each NW was recorded, along with the optical image. This was repeated, moving to a different location on the substrate to identify more wires. The procedure is entirely scaleable, and can be left to run until a suitable NW population is identified, which was 5195 NWs in this study. Each identified NW was then investigated using different experimental modules (spectroscopy and imaging, TCSPC, interferometry and distributed losses), which are discussed in separate sections below.

\subsubsection{Scanning electron microscopy}

SEM measurements were performed on the transferred NWs using the secondary electron detector of a Quanta250 FEG microscope, with an acceleration voltage of \SI{15}{\kilo\electronvolt} and a magnification of 522x. A series of SEM images were measured across a \SI{2}{\milli\meter} by \SI{2}{\milli\meter} square on the transferred NW substrate: each image had a field of view of \SI{290}{\micro\meter} by \SI{250}{\micro\meter}, with a total of 90 images captured.

As the positions of each of the NWs identified in the optical microscopy are defined relative to a series of unique markers, the location (to within $\approx$ \SI{5}{\micro\meter}) of each NW was identified by a transformation (translational and rotational) between the coordinate axes of the optical and SEM images. The machine vision algorithm was applied to identify the same NW in each dataset by identifying the geometrical features. This process enabled the width of the NWs to be determined, beyond the resolution limit of the optical microscopy, whilst permitting the removal of NWs from the analysis that were too close to be observed in the optical. Good registration was achieved for 2492 NWs. More details on the comparison between optical and SEM imaging are given in the SI.

\subsubsection{Spectroscopy and imaging measurements}

Automated PL measurements were performed on single NWs using a quasi-confocal arrangement in the optical microscope. For spectroscopy, the light emitted from an approximately \SI{2}{\micro\meter} diameter spot was collected using an optical fibre. This was connected to a Horiba iHR550 spectrometer to measure the emission spectrum with a resolution of \SI{1}{\nano\meter}. Emission imaging experiments were performed simultaneously by placing a suitable long-pass filter in front of the camera to remove the excitation. 

Lasing experiments used the output from a PHAROS-ORPHEUS \SI{200}{\femto\second} OPA, at a wavelength of \SI{633}{\nano\meter}, as the excitation source, using a band pass filter to clean-up the excitation spectrum. A uniform excitation spot was achieved using a telescopic beam expander and iris, along with a defocusing lens. The result was a top hat excitation spot with a diameter of \SI{73}{\micro\meter}, capable of achieving excitation fluences up to \SI{2000}{\micro\joule\per\centi\meter\squared}. This experiment was performed successfully on 4374 NWs.

PL spectra were fit with a modified Lasher-Stern-Wurfel (LSW) model, that has been adapted from~\cite{Fadaly2020} and \cite{Church2022}, and is given in terms of the photon energy $E$ by~\ref{equ:PL}: 
\begin{equation}
\label{equ:PL}
    I_{LSW}(E) = C E^2 \frac{W(E)} {\text{exp} \frac{E- (E_F + E_g)}{k_B T} - 1},
\end{equation}
where $C$ is an amplitude term, $E_F$ is the quasi-fermi energy, $E_g$ is the energy bandgap of the QW (accounting for confinement energy), $T$ is the carrier temperature and $k_B$ is the Boltzmann constant. $W(E)$ describes the absorption of the NW, which can be approximated using~\ref{equ:absorption}~\cite{Lasher1964,Wurfel1982},
\begin{equation}
\label{equ:absorption}
    W(E) = (1-\text{exp}(B(E))) \left( 1-\frac{2}{ \left( \text{exp} \frac{E- (E_F + E_g)}{k_B T} \right)^{1/2}+1 } \right),
\end{equation}
where $B(E)$ is the 2-dimensional density of states of the QW. Due to disorder in the NW lattice on the nanoscale, the absorption and emission can extend into the expected bandgap with an Urbach tail~\cite{Natsume1995}. To approximate this the model was modified to exponentially decay below the bandgap. Additionally, the spectrum was convoluted with a Gaussian function, $G$, with standard deviation \(\sigma\), which is used to approximate micro-scale symmetrical inhomogeneity in the NW. The resulting fit is described by:
\begin{equation}
\label{equ:PL_fit}
I_{PL}(E)=G(E) \otimes \begin{cases}
    I_{LSW}(E), & \text{if $E>E_{\rm{g}}-dE$}.\\
    A\text{exp}\left(\frac{E-E_{\rm{g}}}{E_{\rm{U}}}\right), & \text{if $E<E_{\rm{g}}-dE$}.
    \end{cases}
\end{equation}
where, $E_U$ is the Urbach energy, \(A\) is an amplitude term and \(dE\) is a small offset energy required to connect the two energy regimes.

The excitation fluence dependent measurements were performed using an automated neutral density filter wheel to change the degree of attenuation of the excitation beam: this provided up to 47 emission spectra per NW, showing both PL and lasing peaks when above the threshold fluence, as demonstrated in Figure~\ref{fig:2}(a). From this data, the threshold fluence was obtained by extrapolating the trend of the lasing peak intensity with fluence. The performance did not depend upon the orientation of the NWs. An example of this analysis is provided in the S.I.

Additionally, the variation of the PL with fluence was extracted by using a median filter to remove the lasing peaks. A global fit of the LSW model was applied to all of the PL spectra, keeping the optical bandgap, inhomogeneous broadening, carrier temperature and Urbach energy constant. This approach determined the quasi-fermi energy, $E_{\text{F}}$, in each QW, as shown in Figure~\ref{fig:2}(b). Importantly, this fit requires two optically active QWs to recover the shape of the emission spectrum of a single NW. To simplify the analysis, the thickness of one of these QWs was fixed to the value obtained in the low fluence PL fitting. Additionally, the $E_{\text{F}}$ in each QW was assumed to be related by a constant. This approach was used to calculate the time-averaged carrier density at every fluence studied, for each optically active QW, $<n_{\text{QW}}>$, using the 2-D density of states, yielding:
\begin{equation}
\label{equ:carr_dens}
    <n_{\text{QW}}> = \frac{\mu E_{\text{F}}}{\pi \hbar^2}  ,
\end{equation}
where $\mu$ is the reduced mass of an electron-hole pair. An example of this calculation is given in Figure~\ref{fig:2}(b). Analysis of the change in PL intensity with excitation fluence also provides information regarding the carrier recombination mechanisms. This data is discussed further in the S.I.

\subsubsection{Time-correlated single photon counting}

The carrier recombination dynamics were measured by routing the optical fibre to a Picoharp HydraHarp400 TCSPC system with Silicon Single Photon Avalanche Photodiodes (SPADs) to measure the PL time decays with a \SI{70}{\pico\second} timing resolution. The PHAROS-ORPHEUS excitation source was used for these experiments with an excitation fluence of approximately \SI{15}{\micro\joule\per\centi\meter\squared}, below the lasing threshold of any of the NWLs. This was performed successfully on 5195 NWs. The experiment was then repeated at \SI{60}{\percent} of the threshold fluence.

An example of the PL decays at low power is shown in Figure~\ref{fig:1}(c). As shown in the S.I. The decay shapes are the result of both radiative and non-radiative decay paths and are well described by a stretched exponential fit, which was used to extract a representative lifetime, $\tau$, given by, 
\begin{equation}
\label{equ:stretched}
    I = I_0 \text{exp}-\left(\frac{t}{\tau} \right)^\beta,
\end{equation}
where $\beta$ is a factor describing the degree of exponentiality, which is an additional measure of disorder in the NWs. $\beta$ has a median value and SD of \SI{0.74(9)}, and correlates with other measures of disorder, such as $\sigma$. More statistics on $\beta$ are provided in the S.I.

The decays close to threshold were used, along with $<n_{\text{QW}}>$, to estimate the initial carrier density in the QWs, $n_{0,\text{QW}}$, at threshold. This was achieved by averaging the area under the PL decay curve, $f(t)$, over the time period of the laser repetition rate (\SI{5}{\micro\second}), using:
\begin{equation}
\label{equ:peak_scale}
    n_{0,\text{QW}} = \frac{\text{\SI{5}{\micro\second}} \times <n_{\text{QW}}>}{\int_{0}^{\text{\SI{5}{\micro\second}}} f(t) \,dt}.
\end{equation}
This approach enabled the calculation of the threshold carrier density - which is a more direct measurement of the lasing threshold than the threshold fluence, as it decouples the lasing mechanism from complicated effects such as light absorption in the NW and carrier capture in the QWs. The S.I contains more details regarding the PL decays close to threshold, along with a comparison between the threshold fluence and threshold carrier densities.

\subsubsection{Interferometric measurements of cavity reflectivity}

To measure the reflectivity of the NW laser cavity, the NWs were excited with a fluence above the lasing threshold. The emission from the NW end-facets was analysed by a time-gated Michelson interferometer that has been previously discussed~\cite{Skalsky2020}. For each NW, one of the interferometric mirrors was automatically translated through the zero path length position, using SPADs to detect the interference fringes, such as those inset in Fig~\ref{fig:2}(c). To accelerate the experiment, the interference fringes were sampled at different positions along the path. These measurements were successfully performed for 724 NWs.

The interferograms for each NW were automatically processed using a Fourier filter to extract the interference fringes associated with frequencies close to the lasing wavelength. A peak detection algorithm was applied to calculate the fringe visibility, $V$, from the peak intensities $I$, at different distances, defined by:
\begin{equation}
\label{equ:visibility}
    V = \frac{I_{\text{max}} - I_{\text{min}}}{I_{\text{max}} + I_{\text{min}}}.
\end{equation}
An example of the fringe visibilities are shown in Figure~\ref{fig:2}(c). The coherence length, $L_{\text{coh}}$ of the lasing light was extracted from a FWHM of a Gaussian fit to the data. This value was then used to calculate the cavity reflectivity, R, using:
\begin{equation}
\label{equ:reflect}
    R = \text{exp}\left(\frac{-2 \pi n L}{L_{\text{coh}}}\right),
\end{equation}
where $L$ is the NW length, extracted from the microscopy. $n$ is the modal refractive index of the wire. As the NW widths are large, and the lasing modes are well confined to the NW, we approximate this to be equal to the bulk value of 3.4~\cite{Adachi1989}. The refractive index also varies slightly between different NWs: by analysing the longitudinal mode spacing, the upper limit of the uncertainty on this parameter was estimated to be +/-0.5, or a percentage uncertainty of \SI{15}{\percent} (more information can be found in the S.I.). $R$ is the geometric mean of the reflectivity of both NW end-facets. The reflectivity of each end-facet was extracted using complementary imaging, which is discussed in the S.I.

\subsubsection{Distributed losses}

The distributed losses were measured by exciting the NWs using a \SI{2}{\micro\meter}-diameter focused spot from a low-power CW HeNe laser, and examining the spatial distribution of the low fluence emission using the imaging camera. The diffusion length in the NWs is expected to be sub-\SI{}{\micro\meter}, and so carriers recombine close to where they are generated. As shown in Figure~\ref{fig:2}(d), this results in a bright emission from the NWs that is co-incident with the excitation spot. However, a portion of the emitted light is confined to the NW due to waveguiding effects: this light travels along the wire and couples out of the end-facets, which is seen in the image.

To quantify the attenuation of the light as it travels down the wire, the automated microscopy approach was used to find the lengths and in-plane orientation angles of each nanowire: the excitation spot position was then automatically scanned along the length of the wire, and the intensity of light coupling from each facet was monitored by recording a series of images. This approach is similar to that employed in ref~\cite{Barrelet2004}, and was applied to 267 NWs, an example is shown in Figure~\ref{fig:2}(d).  

This data was fit using Equation~\ref{equ:distributed}, where the intensity ratio is studied to minimize the impact of changes in the coupling of the excitation into the NW due to slight misalignment during the scan.

\medskip
\textbf{Supporting Information} \par 
Supporting Information is available from the Wiley Online Library or from the author.

\medskip
\textbf{Acknowledgements} \par This was work funded by UKRI under grant MR/T021519/1 and EP/V036343/1. Research data supporting this publication will be made available at DOI: 10.48420/21865785, and the code to perform the analysis will be made available at DOI: 10.48420/21865545, and on github: https://github.com/p-parkinson.

CRediT author statement: \textbf{Stephen Church}: Formal analysis, Investigation, Methodology, Software, Visualisation, Writing - original draft. \textbf{Nikesh Patel}: Formal analysis, Investigation, Writing - review and editing. \textbf{Ruqaiya Al-Abri}: Investigation, Writing - review and editing. \textbf{Nawal Al-Amairi}: Formal analysis, Writing - review and editing. \textbf{Yunyan Zhang}: Resources, Writing - review and editing. \textbf{Huiyun Liu}: Resources, Writing - review and editing. \textbf{Patrick Parkinson}: Conceptualization, Data curation, Funding acquisition, Methodology, Software, Supervision, Writing - review and editing.

\providecommand{\latin}[1]{#1}
\makeatletter
\providecommand{\doi}
  {\begingroup\let\do\@makeother\dospecials
  \catcode`\{=1 \catcode`\}=2 \doi@aux}
\providecommand{\doi@aux}[1]{\endgroup\texttt{#1}}
\makeatother
\providecommand*\mcitethebibliography{\thebibliography}
\csname @ifundefined\endcsname{endmcitethebibliography}
  {\let\endmcitethebibliography\endthebibliography}{}


\renewcommand{\thefigure}{S\arabic{figure}}
\renewcommand{\thetable}{S\arabic{table}}
\renewcommand{\theequation}{S\arabic{equation}}
\setcounter{figure}{0} 
\setcounter{equation}{0} 
\setcounter{table}{0} 

\newpage
\section{Supporting Information}

\section{Nanowire dimension statistics}

The dimensions of each NW were determined using different experimental approaches. Firstly by the automated dark field optical microscopy approach. From these measurements, the median width (and standard deviation) was found to be \SI{2.1(4)}{\micro\meter} and the length was found to be \SI{14.6(5)}{\micro\meter}. While the distribution of widths is largely symmetric, the length distribution is strongly skewed towards the maximum length. A series of SEM images were obtained across the NW population that enabled a second, higher resolution, measurement of the NW dimensions to be performed. These SEM results produced NW widths of \SI{0.7(1)}{\micro\meter}. As the SEM measurements have an improved resolution, these values are utilised as the NW width for further analysis.

\begin{figure*}
    \centering
    \includegraphics[width = 0.95\linewidth]{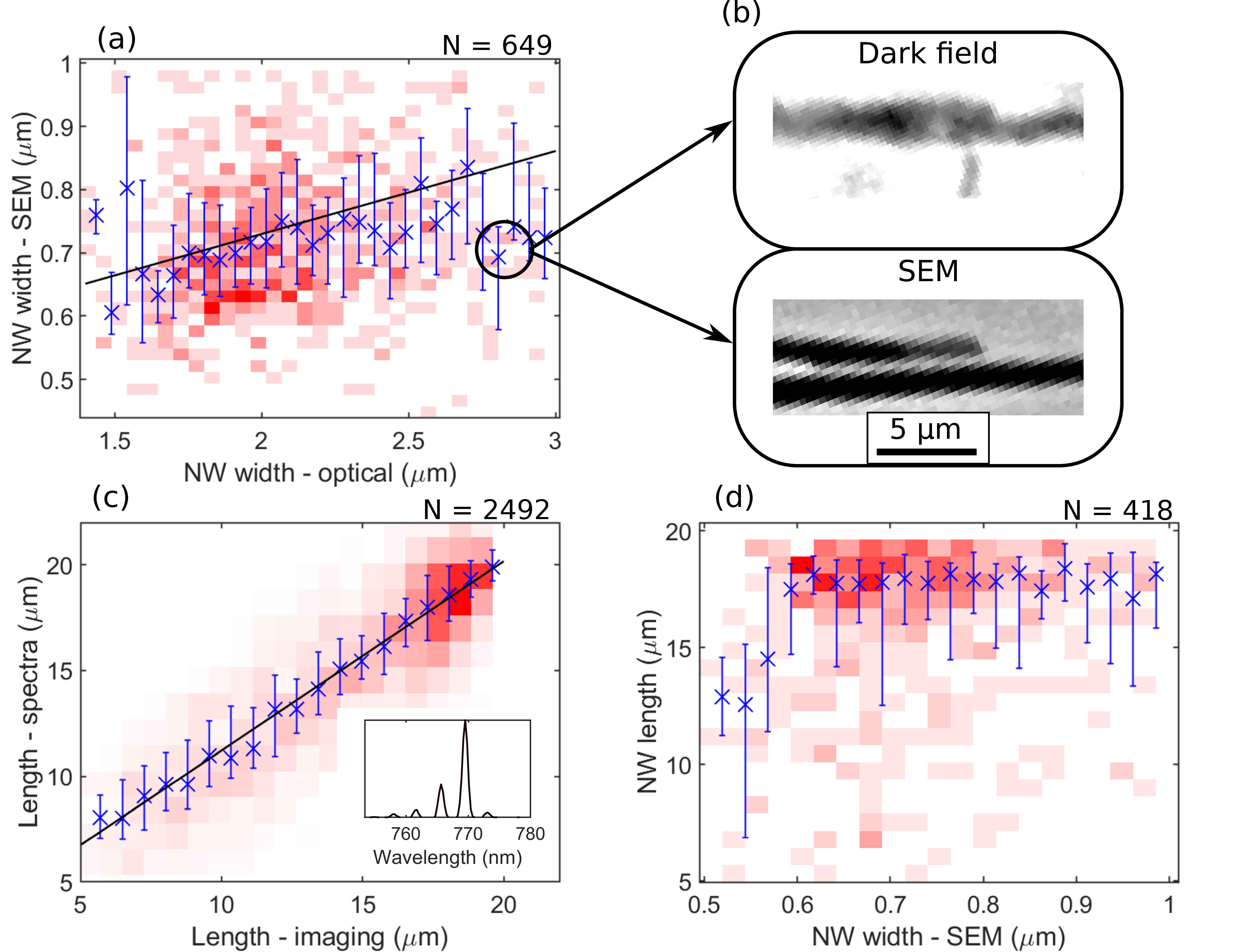}
    \caption{Statistics on the NW geometries, measured using different experiments. In these graphs, the blue points repesent the median and inter-quartile range of the values in each x-axis bin, and the black line is a linear fit to the dataset. (a) A two dimensional histogram showing the correlation between measurements of the NW width taken using dark field optical microscopy and SEM. (b) A comparison of the imaging of two closely-spaced NWs using dark field optical microscopy and SEM. (c) A two dimensional histogram showing a comparison of the NW length, obtained from dark field optical imaging and from analysis of the lasing longitudinal mode spacing. Inset is an emission spectrum demonstrating peaks from each longitudinal mode. (d) A two dimensional histogram showing a comparison between the NW length and width; no correlation is observed.}
    \label{fig:S1}
\end{figure*}

Fig.~\ref{fig:S1} shows that the two width measurements are correlated for narrow NWs, however, for optical widths greater than \SI{2.5}{\micro\meter}, the correlation is not present. Inspection of the images demonstrates that the reduction in correlation is due to the presence of NWs which are spaced by $<$\SI{1}{\micro\meter}: these multiple NWs are erroneously identified as a single NW in the optical microscopy. An example of this comparison is shown in Fig.~\ref{fig:S1}(b), and these NWs were removed from subsequent analysis.

The NW length was also be extracted by measuring the spectral spacing of the longitudinal modes of the Fabry-P\'erot cavity. These modes were observed in the emission spectra of the NWs at a fluence above the lasing threshold, an example is shown inset in Fig.~\ref{fig:S1}(c). Using the refractive index of the material, $n$, taken to be 3.4~\cite{Adachi1989}, the length, $L$, was calculated using:
\begin{equation}
    \label{equ:length}
    L = \frac{hc}{2n\Delta E}
\end{equation}
where, $\Delta E$ is the mode spacing. This analysis was performed for the lasing NWs, and the calculated length was found to be consistent with the observed length from imaging for 2492 of the NWs, and resulted in a strong positive correlation between the two results, as shown in Fig.~\ref{fig:S1}(c). 

For the analysis in this report we have approximated the effective refractive index to have a constant value for all nanowires. If it is assumed that the spread in the length data is entirely due to variation in the modal refractive index we can use Figure S1(c) to estimate an upper limit on the uncertainty on this parameter. This gives a value of n = \SI{3.4(5)}, or a percentage error of \SI{15}{\percent}. This uncertainty is small compared with the large spread of reflectivity values shown in Figure 3(d), and thus justifies this approach.

This approach also allows the calculation of the refractive index of each individual NW. There is no correlation between refractive index and NW width – supporting the argument that the lasing modes are well confined to the NW cavity.

This analysis failed on NWs that did not have a consistent mode spacing: this primarily occurred due to the detection of lasing emission from nearby NWs. For the analysis of the impact of the NW cavity, including distributed losses and reflectivity, only those NWs where this analysis was successful were included.

The result of this multi-modal analysis is a population of NWs that have consistent results for NW dimensions. Fig.~\ref{fig:S1}(d) shows the resultant distributions of these parameters. There is no significant correlation between the NW length and width. This is expected for these samples, because the NW length is dictated by the duration of the VLS NW growth, whereas the NW width is determined by the subsequent growth of the core/shell structure.

\section{Statistics on the quantum well disorder}

The PL spectrum from each NW was fit with a PL lineshape that is adapted from the LSW model\cite{Fadaly2020,Church2022}, and is defined by equations (3), (4) and (5) in the main paper. These equations contain two parameters that characterise the disorder in the samples. $\sigma$ is the standard deviation of a Gaussian that is convoluted with the PL lineshape: $\sigma$ is therefore a measure of the symmetrical disorder in the QWs, which may be due to, for example, variation in the QW width, the strain in the QWs and the composition of the barriers within each NW. $E_U$ is the Urbach energy, which represents a modification of the density of states of the QWs close to the bandgap, due to disorder on the atomic scale~\cite{Natsume1995}, and is therefore an asymmetrical measure of the disorder.

\begin{figure*}
    \centering
    \includegraphics[width = 0.95\linewidth]{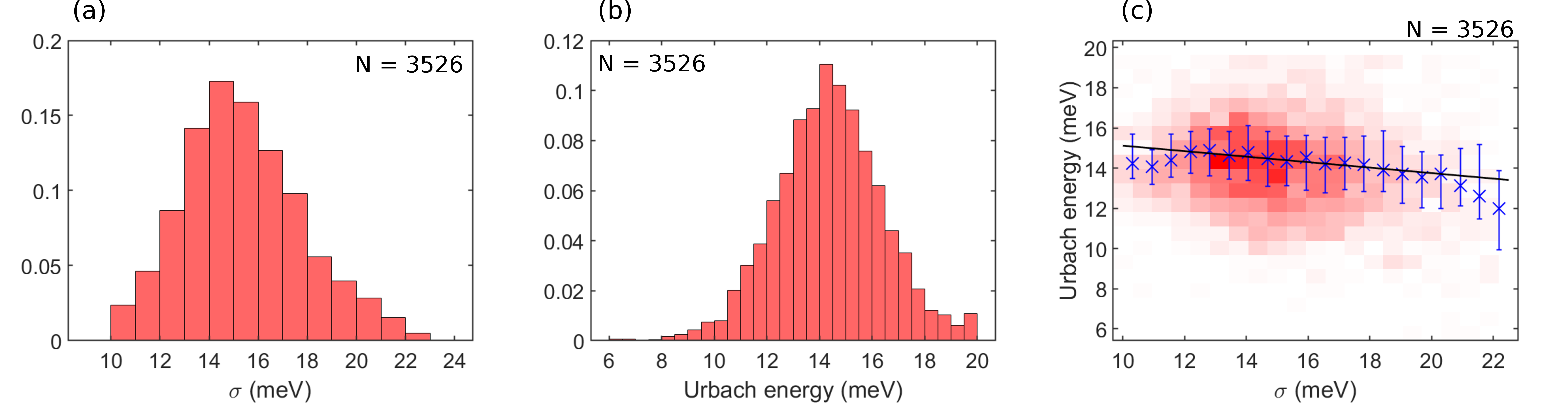}
    \caption{Graphs showing the distribution of disorder parameters in the QWs, extracted from fits to the lineshape of the low power PL spectra. (a) A histogram showing the distribution of the $\sigma$ parameter, extracted from the width of the PL spectra. (b) A histogram showing the distribution of the Urbach energy, extracted from the low energy region of the PL spectra. (c) A two dimensional histogram showing the relationship between the $\sigma$ parameter and the Urbach energy. The blue points represent the median and inter-quartile range of the values in each x-axis bin, and the black line is a linear fit to the dataset.}
    
    \label{fig:S2}
\end{figure*}

$\sigma$, for the active lasing QW, was extracted for 3526 NWs, providing the histogram shown in Fig.~\ref{fig:S2}(a): this has a median (and standard deviation) value of \SI{15(2)}{\milli\electronvolt}, with a range of values between \SI{10}{\milli\electronvolt} and \SI{23}{\milli\electronvolt}. These values are lower than those obtained from studies of AlGaAs/GaAs core/shell QW NW systems~\cite{Alanis2017}, which suggests that these QWs are of improved uniformity.

$E_U$ was extracted for 3526 lasing NWs, and the distribution is shown in Fig.~\ref{fig:S2}(b). The median value is \SI{14(2)}{\milli\electronvolt}, with a range between \SI{8}{\milli\electronvolt} and \SI{20}{\milli\electronvolt}.

Fig.~\ref{fig:S2}(c) demonstrates the relationship between these two disorder parameters, this has a Pearson's $r$-parameter of -0.16 and a $p$-value of close to 0. This weak relationship suggests that these two measurements of disorder are coupled. Additionally, $E_U$ does not correlate strongly with any of the lasing metrics - indicating that this inhomogeneous broadening is not a key factor in limiting the NWL performance.

\section{Carrier dynamics}    
    
\begin{figure*}
    \centering
    \includegraphics[width = 0.95\linewidth]{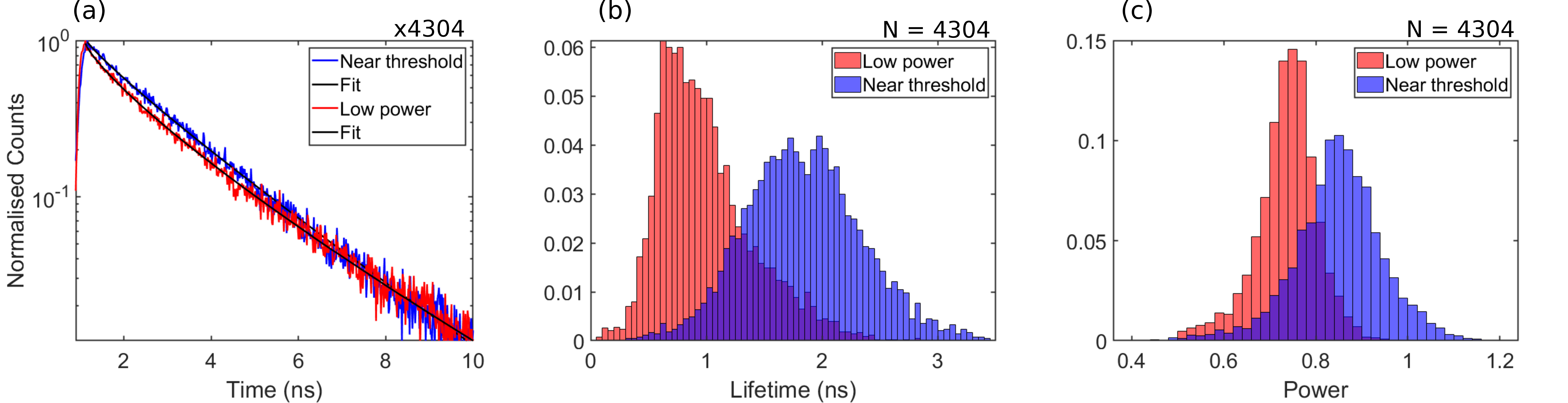}
    \caption{Measurements of the PL recombination dynamics at low power, and close to threshold (at 0.6 P\textsubscript{th}). (a) An example decay curve at both powers for the same NW, with a stretched exponential fit applied to each dataset. (b) Histograms showing the variation in recombination lifetime across the population of NWs at both excitation powers. (c) Histograms showing the variation in power factor across the population of NWs at both excitation powers.}
    \label{fig:S3}
\end{figure*}

As discussed in the main paper, PL decays were measured close to the lasing threshold (at 0.6 P\textsubscript{th}), in order to determine the carrier densities generated by optical excitation. These measurements resulted in, on average, a longer recombination lifetime and a different decay shape, when compared to low power measurements. An example of these decays on the same NW is shown in Fig.~\ref{fig:S3}(a); a stretched exponential fit was applied to each curve. The near threshold measurement demonstrates a lifetime of \SI{2.3}{\nano\second} and a power factor of 0.9, compared with \SI{1.4}{\nano\second} and 0.8 for the low power measurement. The increased lifetime at the higher power may be a consequence of saturation of non-radiative defects.

This measurement and analysis was performed for 4304 NWs; the recombination lifetimes for the NW population are shown in Fig.~\ref{fig:S3}(b). The median lifetime increases from \SI{0.9(4)}{\nano\second} to \SI{1.8(5)}{\nano\second} near threshold, likely due to saturation effects. These two lifetimes are strongly correlated, with a Pearson's r coefficient and p-value of 0.4 and close to 0; this suggests that the fundamental recombination mechanisms are unchanged at the elevated powers. The power factors across the population are shown in Fig.~\ref{fig:S3}(c), with a median value of 0.74(7), increasing to 0.85(9) near threshold. The power factors at each power are also positive correlated with an r coefficient and p-value of 0.14 and close to 0. A portion (\SI{5}{\percent}) of these power factors are greater than 1, however the fitting error on these parameters are approximately \SI{5}{\percent}, the vast majority of these values are consistent with 1 at the $3\sigma$ level, and are therefore mono-exponential.

\section{Analysis of light-in light-out curves}    

The fluence-dependent high-throughput measurement was used to extract the lasing threshold fluence and the lasing wavelength at threshold. This was achieved using a simple, and computationally fast, algorithm that was applied to the spectra taken at all measured fluences. This algorithm was applied during the main experimental loop, and therefore analysed the emission spectrum of the NW on-the-fly. The lasing threshold was therefore identified during the experiment, enabling the measurements on each NW to be terminated after collecting the useful data.

\begin{figure*}
    \centering
    \includegraphics[width = 0.95\linewidth]{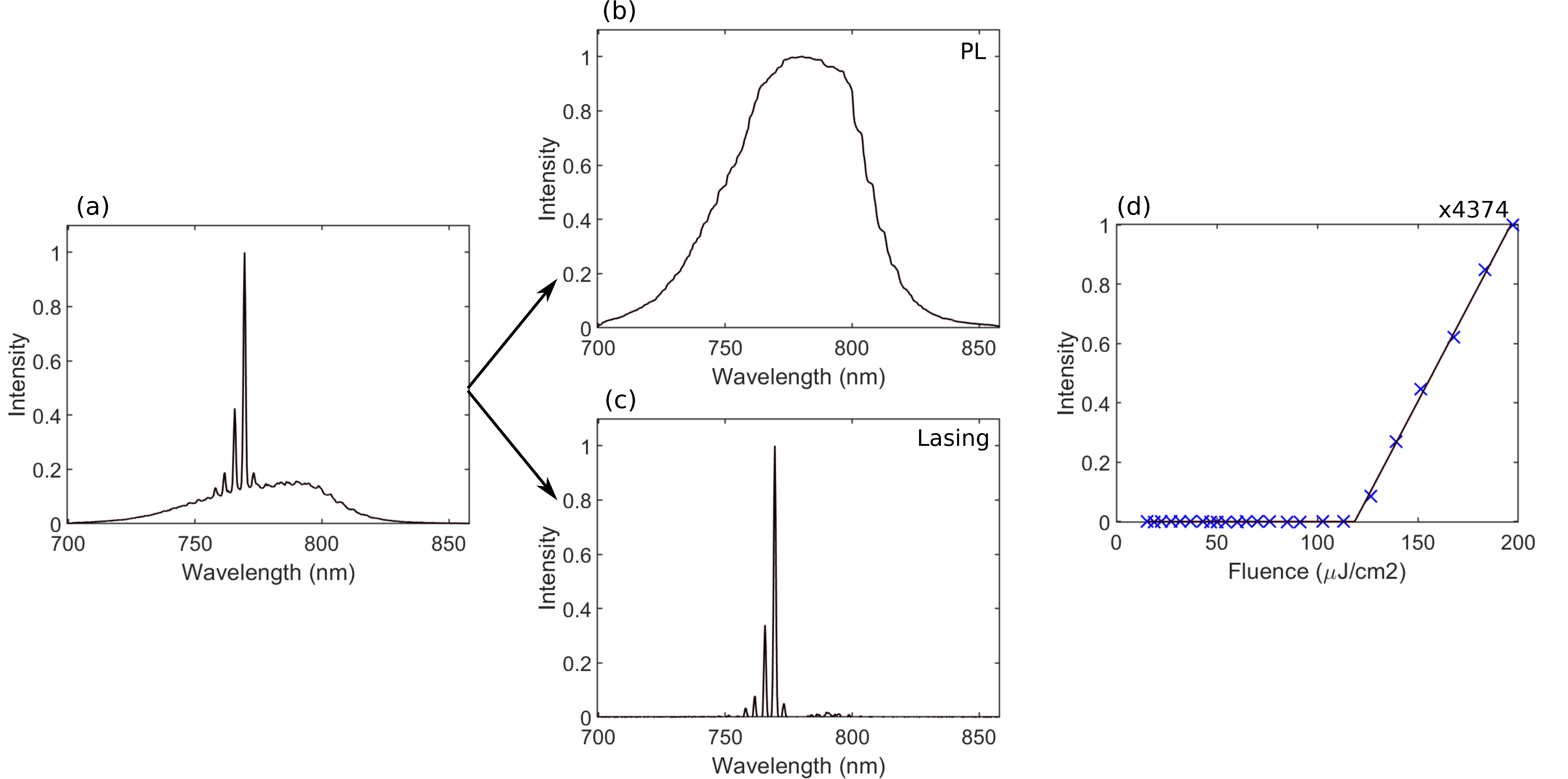}
    \caption{A demonstration of the analysis to determine the lasing threshold of each NWL. (a) An unprocessed emission spectrum from a NWL, excited above the lasing threshold, showing PL and lasing peaks. (b) The processed spectrum, using a median filter to remove the lasing peaks, leaving only the PL peak. (c) The result of subtracting (b) from (a), to isolate the lasing peaks. (d) A light-in-light-out curve, showing the dependence of the intensity of the extracted lasing peaks with excitation fluence. }
    \label{fig:S4}
\end{figure*}

An illustration of this algorithm is shown in Fig.~\ref{fig:S4}. The measured emission spectrum (Fig.~\ref{fig:S4}(a)) was processed by applying a median filter to isolate the PL emission peak (Fig.~\ref{fig:S4}(b)): this was subsequently subtracted from the original spectrum to isolate any lasing peaks (Fig.~\ref{fig:S4}(c)). The processed spectrum was then passed through multiple checks to determine if lasing was observed: the peak intensity must be greater than a threshold value, and significantly greater than the median intensity. Additionally, the full-width at half maximum (FWHM) must be smaller than a threshold value (typically \SI{3}{\nano\meter}). Finally, the peak must consist of at least two points in the spectrum, to avoid misidentifying cosmic rays. Once this algorithm detected lasing, only a few more spectra at elevated fluences were recorded - to produce a suitable light-in light-out curve (LILO), an example of which is shown in Fig.~\ref{fig:S4}(d).

The LILO curve was fit using a linear model for the intensity, $I$, with a threshold fluence, $P_{\rm{th}}$, defined by: 
\begin{equation}
\label{equ:LILO}
  \tau(I)=\begin{cases}
    A, & \text{if $P<P_{{\rm{th}}}$}.\\
    A + m(P- P_{{\rm{th}}}), & \text{if $P>P_{{\rm{th}}}$}
  \end{cases}
\end{equation}
where \(m\) is the slope of the variation, and \(A\) is a small offset. 

This approach does not reach the fluences required to observe the commonly observed "s-shape" LILO curve that is characteristic of NW lasers~\cite{Gao2013}, and therefore the extracted lasing thresholds are likely to be less accurate than a fit to this curve. However, the approach in this study is sufficient to study the variation in threshold and also significantly reduces the time required for the full experimental loop, whilst preventing degradation of the NWs at excitation fluences far above threshold.
    
\section{Comparison between threshold fluence and threshold carrier density}

Fig.~\ref{fig:S5}(a) compares the threshold carrier densities with the threshold fluence. A strong, linear, positive correlation is observed between these two, independently determined, parameters, with r and p-values of 0.30 and close to 0. This correlation validates the approach to extract the carrier densities, as increasing the excitation density directly leads to an increase in the carrier density. 

However, Fig.~\ref{fig:S5}(a) demonstrates that there is a degree of scatter on the points. This scatter is likely related to differences in the structure of each NW: which will have an influence on both the light coupling into the NW and the absorption in the QWs. To provide an insight into these effects, COMSOL simulations were performed to investigate the coupling between incident light and the NWs. These simulations followed a similar procedure to that reported in~\citen{Church2022}. 

An example simulation is shown in Fig.~\ref{fig:S5}(b), which consists of a GaAsP NW in air sitting on a Si substrate, ignoring absorption effects. Light at a wavelength of \SI{635}{\nano\meter} is incident from the top of the diagram. The simulations demonstrate an expected increase in wavelength as the light propagates through the Si layer and in the NW. The electric field is enhanced at the apexes of the NW cross section and at the substrate surface~\cite{Wells2012}.

The light coupling into the NW was simulated on a series of NWs with different widths. This was achieved by assuming that the NW is strongly absorbing: the excitation photon energy was therefore above the bandgap, and all light that coupled into the NW was absorbed. These results are shown in Fig.~\ref{fig:S5}(c), for a uniform spot of diameter \SI{73}{\micro\meter} - as used in the experiment.

The fraction of light that couples into the NWs increases approximately linearly with NW width. In this study, the NW width has a median value of \SI{0.7}{\micro\meter}: therefore approximately \SI{0.75}{\percent} of the incident photons will couple into the NW. Normalising this value to the size of the NW shows that this is approximately \SI{80}{\percent} of the light that is directly incident on the wire: this reduction is mainly due to reflection from the air/NW interface.

For NWs with a width greater than the median, the NW width is larger than the excitation wavelength, and the normalised coupling efficiency remains constant. In this regime, the photon density in the NW can be considered, in a first order approximation, to be independent of the NW width. 

However, for NWs narrower than the median, the NW width is comparable, or smaller than the excitation wavelength, which causes a reduction in the normalised coupling efficiency down to \SI{50}{\percent} for a width of \SI{0.2}{\micro\meter}. The exception to this trend are \SI{0.5}{\micro\meter} and \SI{0.4}{\micro\meter} NW, that demonstrate a resonant enhancement in the coupling strength: which may be of interest to explore further. In summary, for the NWs in this study, variation in the coupling efficiency will impact the energy density in the NW and have an influence in the thresholds.

In the experiments in this study, the excitation wavelength was chosen to only excite the QWs. To investigate the variation in absorption between the QWs, a test structure was simulated, shown in Fig.~1(d) in ref~\citen{Skalsky2020}. Here, the NW width was \SI{300}{\nano\meter} and three QW layers were considered, with widths of \SI{7.5}{\nano\meter}, \SI{3.5}{\nano\meter} and \SI{6}{\nano\meter}, at different distances from the NW core. 

The simulated QWs had relative cross sectional areas of 0.3, 0.4 and 1 respectively. It was calculated that the proportion of the incident light absorbed in each layer was \SI{0.6}{\percent}, \SI{0.5}{\percent} and \SI{0.8}{\percent} respectively. There was therefore no strong correlation between the QW cross sectional area and the absorption, with a factor of 3 increase in the area only corresponding to a 0.3x increase in the absorption. 

Consequently, these simulations suggest that the spatial overlap between the QWs and the electric fields that plays a more important role in the absorption, than simply the volume of each QW. However, this simple model does not account for inhomogeneities in the NW core-shell structure and therefore cannot be used to accurately calculate the absorption in each QW for comparison to the experiment.

\section{Fluence dependence of the photoluminescence spectrum}    
    
\begin{figure*}
    \centering
    \includegraphics[width = 0.95\linewidth]{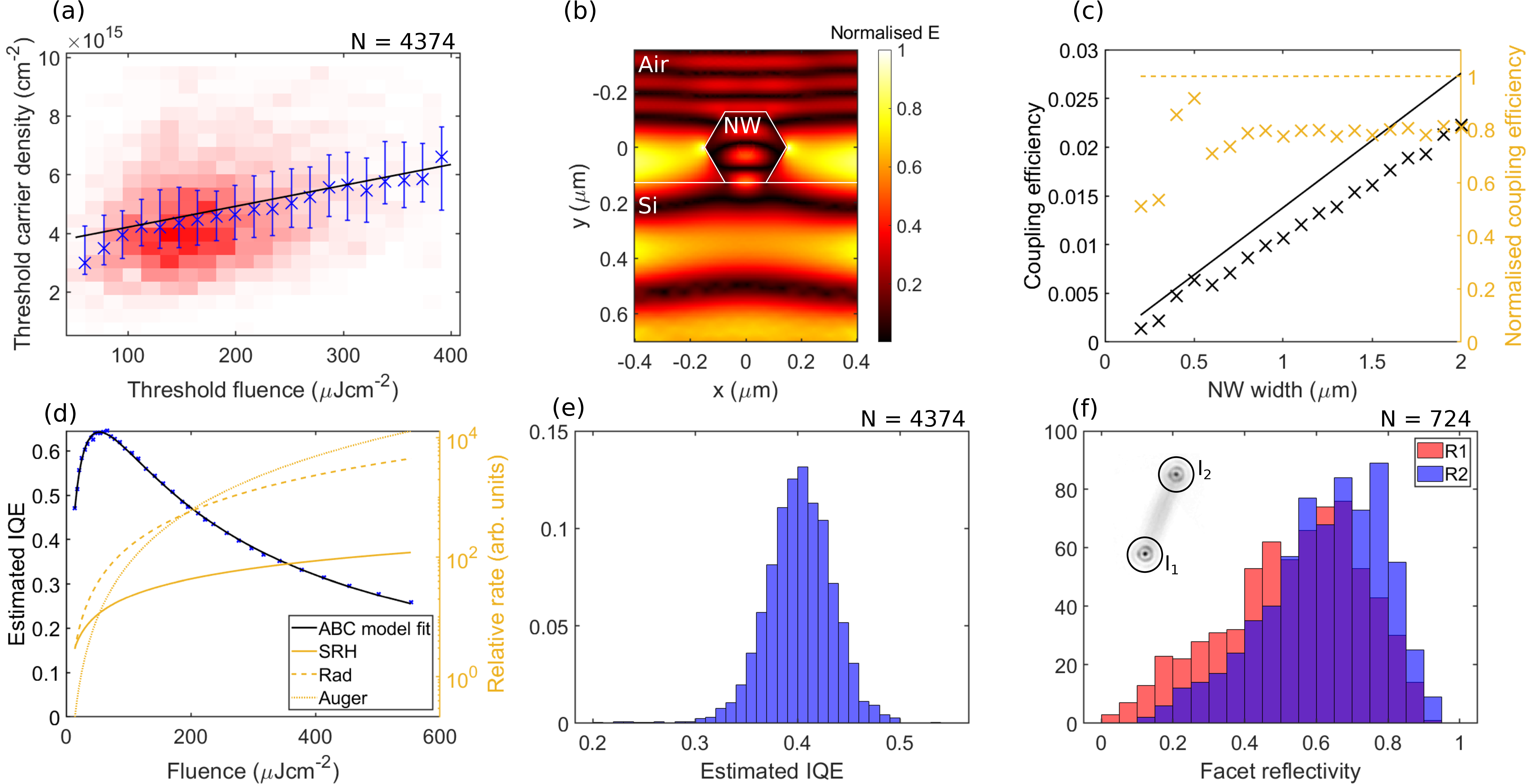}
    \caption{(a) A two dimensional histogram showing a positive correlation between the independently determined threshold fluence and threshold carrier density. (b) An example output from a COMSOL simulation of light coupling into a NW of \SI{0.25}{\micro\meter}, showing the normalised magnitude of the electric field, and ignoring absorption effects. (c) The output of COMSOL simulations, showing the variation of the light coupling efficiency into the NWs with changing NW width, and the coupling efficiency normalised to the NW width. The straight lines represent a perfectly absorbing NW. (e) The variation in the estimate IQE of the PL emission, taken from the integrated intensity, with changing excitation fluence. Below the lasing threshold, the data has been fit with an ABC model of recombination, which extracts the relative rates of SRH, radiative and Auger recombination. (e) A histogram showing the estimated IQE at the lowest fluence, extracted from an ABC model fit to (d). (f) A histogram showing the relative reflectivity of each NW facet. Inset is an image of a lasing NW, which demonstrates how the reflectivity of each facet is isolated by image analysis. }
    \label{fig:S5}
\end{figure*}

The analysis approach shown in Fig.~\ref{fig:S4} was used to isolate the PL from the emission spectrum from each NW at each excitation fluence. The fluence-dependence of the PL intensity was then studied to provide further insights into the carrier recombination mechanisms in the QWs. An example of this analysis is shown in Fig.~\ref{fig:S5}(d), which plots the estimated IQE, proportional to the PL intensity divided by the excitation fluence. The efficiency increases superlinearly with fluence and reaches a peak efficiency at \SI{55}{\micro\joule\per\centi\meter\squared}; above this fluence, the efficiency drops rapidly. 
    
As the peak efficiency occurs at fluences far below the lasing threshold, the efficiency trends are likely to be due to the impact of different recombination mechanisms in the material, rather than a depletion of the spontaneous emission due to pumping of the stimulated emission. The fluence dependence of the efficiency, $I(p)/p$ was fit using an ABC model~\cite{Karpov2014},
\begin{equation}
\label{equ:IQE}
    \frac{I(p)}{p} = D \frac{Bp^2}{Ap + Bp^2 + Cp^3},
\end{equation}
where $p$ is the excitation fluence, $Ap$ is the Shockley-Read-Hall (SRH) rate, $Bp^2$ is the radiative rate, $Cp^3$ is the rate of Auger recombination, and $D$ is a scaling constant. As shown in Fig.~\ref{fig:S5}(d), this equation fits well across the whole range of fluences considered. This allows the extraction of each rate; the IQE of the PL emission was also estimated by considering Equation~\ref{equ:IQE}, without the factor $D$. At low fluences, Auger recombination is negligible and SRH and radiative recombination are comparable, the IQE is \SI{47}{\percent}. As the fluence increases, SRH becomes less important, and the IQE peaks at \SI{65}{\percent}. Auger recombination is the fastest process at the highest fluences studied, and the IQE is therefore the minimum at \SI{23}{\percent}. 

This analysis was performed for 4374 NWs, the IQE at low fluences for this populations is shown in Fig.~\ref{fig:S5}(e). This IQE has a median value of 0.40(5). As the median low-fluence lifetime has been measured to be \SI{0.74(6)}{\nano\second}, the median radiative lifetime was therefore calculated to be \SI{1.9}{\nano\second} and the non-radiative lifetime was \SI{1.2}{\nano\second}.
    
\section{Reflectivity of individual end facets}   

As discussed in the main paper, the reflectivity of the Fabry-P\'erot cavity, was calculated for each NW from measurements of the coherence length, and utilising Equation (10). This is an intensity reflectivity, and is effectively a geometrical average of the geometries of each of the cavity end facets. 

As shown in Fig.~2(d), it is possible to detect light coupling out of each end facet using far-field imaging. Similar images were recorded when exciting above the lasing threshold, an example of these is shown inset in Fig.~\ref{fig:S5}(f). The intensity of light coupling from one of the facets is higher than the other. This reflects a difference in facet transmission and therefore a difference in facet reflectivity.

The intensity of light coupling through each facet was measured as the sum of the pixel intensities in a circle centred on each facet. These intensities were used to calculate the reflectivity of each facet $R_{1,2}$, from the cavity reflectivity $R$:
\begin{align}
    \label{equ:ref}
    R^2 &= R_1 R_2 \\
    R_{1,2} &= 1 - \eta I_{1,2}
\end{align}
where $I_{1,2}$ is the intensity of light emitted from each facet and $\eta$ is a normalisation factor that accounts for transmission and scattering effects. This analysis was performed for 4374 NWs: a histogram demonstrating the distribution of $R_1$ and $R_2$ is shown in Fig.~\ref{fig:S5}(f). The lower reflectivity facet has a median reflectivity of 0.55(19) and the other facet has 0.64(17). We propose that the least reflective facet is from the tip of the NW, due to defects that form at this interface\cite{Zhang2019a}. In contrast, the reflectivity of the bottom facet is likely to be high due to cleaving of his facet during the NW transfer process~\cite{Alanis2019a}.

\section{Calculations of the quantum well widths}  

    \begin{figure*}
    \centering
    \includegraphics[width = 0.95\linewidth]{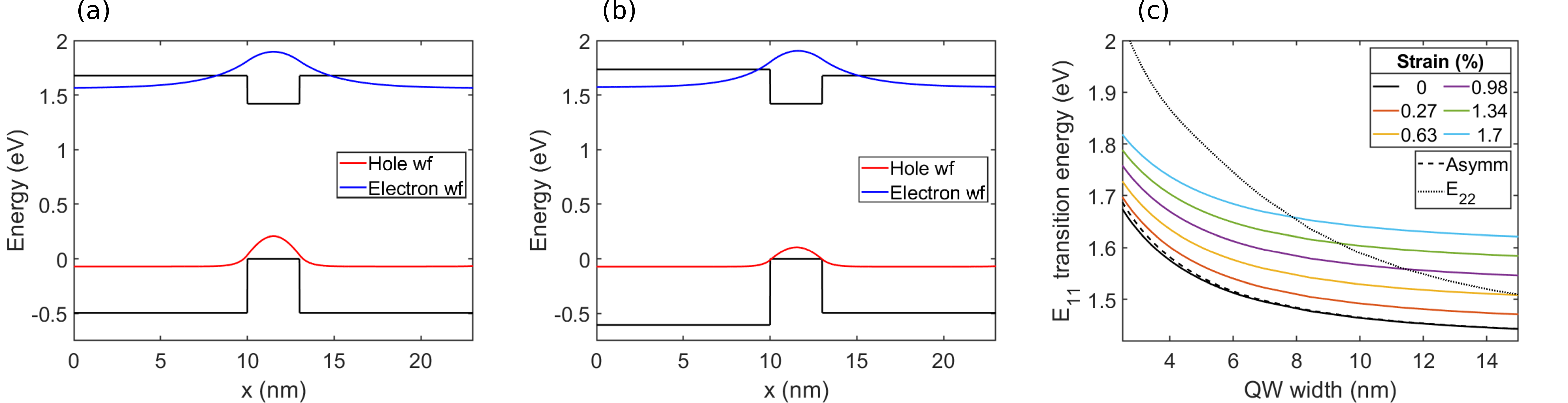}
    \caption{(a) The band structure of a \SI{3}{\nano\meter}-thick GaAs/GaAsP QW, assuming no strain, and the ground state wavefunctions of electrons and holes. (b) The band structure of the asymmetrical inner core GaAs/GaAsP QW, with a \SI{3}{\nano\meter} thickness and ground state wavefunctions. (c) The calculated ground state $E_{11}$ transition energies for symmetrical QWs with different widths, assuming different degrees of internal strain. Also included are the $E_{11}$ transition energies for an symmetrical QW with zero strain, and the $E_{22}$ transitions energies of the symmetrical QW with no strain.   }
    \label{fig:S6}
\end{figure*}

The widths of the QWs were derived from the fits to the PL emission peaks. These fits yielded the PL bandgap, which were used to determine the QW widths by means of a look-up table that was obtained by calculating the confined states of electrons and holes in QWs with different widths.

The QW bandstructure for the GaAs/GaAs QWs was extracted from previous modelling~\cite{Skalsky2020}. Two situations were considered: Fig~\ref{fig:S6}(a) shows the case of one of the outer two QWs, which have symmetrical barriers, whilst Fig~\ref{fig:S6}(b) shows an example of the inner core-QW with asymmetrical barriers. In the core-shell structure, the QWs are separated by \SI{20}{\nano\meter} and so can be considered in isolation.

The energy levels and wavefunctions for electrons and holes were calculated using Numerov's method~\cite{Allison1970}, assuming a hole effective mass of $0.34m_e$~\cite{Molenkamp1988} and an electron effective mass of $0.063m_e$~\cite{Levinshtein1996}. In this approach, continuous solutions to the Schr\"odinger equation are calculated using an established formula~\cite{Allison1970}: the bound states are found by iterating on the state energy until the appropriate boundary conditions are met, in this case the wavefunction must tend to zero far from the QWs. This approach has been demonstrated to be a rapid way of calculating bound energy states~\cite{Allison1970}. This calculation was repeated for QWs of different widths, as shown in Fig.~\ref{fig:S6}(c), and the transition energy drops with increasing width, as is typical for a QW system.

As discussed in previous studies~\cite{Zhang2019a}, the lattice mismatch between the barriers and the QWs is expected to lead to \SI{1.7}{\percent} compressive strain in the QWs. This strain will increase the QW bandgap and lead to higher optical transition energies; this leads to $E_{11}$ transition energies between \SI{1.82}{\electronvolt} and \SI{1.62}{\electronvolt} with increasing width. However, these calculated energies are higher than all of the experimentally measured PL bandgaps (\SI{1.55}{\electronvolt} and \SI{1.60}{\electronvolt} for the two emission peaks). This result suggests that the strain is, at least partially, compensated in the QWs.

To quantify the impact of strain, the same calculations were repeated for degrees of QW strain between \SI{0}{\percent} and \SI{1.7}{\percent}, these results are shown in Fig.~\ref{fig:S6}(c). As the internal strain is reduced, $E_{11}$ drops by up to \SI{170}{\milli\electronvolt}, having a larger effect for wider NWs. For strains below \SI{1}{\percent}, $E_{11}$ overlaps with the measured energies and therefore an estimated QW width can be determined. Therefore, the QW widths reported in this study are the "zero-strain equivalent widths". As the strain is likely to be non-zero in reality, the true QW widths will be larger than these values, however, this will only impact the magnitude of the QW widths and not any of the observed trends, and this will therefore not effect any of the major results.

The Numerov method was also used to calculate the excited states in each QW, which was used to determine the $E_{22}$ transitions, which are shown in Fig.~\ref{fig:S6}(c) for a zero strain QW. The $E_{22}$ transition is at between \SI{60}{\milli\electronvolt} and \SI{350}{\milli\electronvolt} higher energy than the $E_{11}$ transition, with a larger difference for narrower QWs. In comparison, the energies of the $E_{11}$ transitions in the asymmetrical QW are also shown in Fig.~\ref{fig:S6}(c). In this case, the energy splitting has a maximum value of \SI{16}{\milli\electronvolt} for narrow QWs.

\medskip
\providecommand{\latin}[1]{#1}
\makeatletter
\providecommand{\doi}
  {\begingroup\let\do\@makeother\dospecials
  \catcode`\{=1 \catcode`\}=2 \doi@aux}
\providecommand{\doi@aux}[1]{\endgroup\texttt{#1}}
\makeatother
\providecommand*\mcitethebibliography{\thebibliography}
\csname @ifundefined\endcsname{endmcitethebibliography}
  {\let\endmcitethebibliography\endthebibliography}{}

\end{document}